\documentclass[12pt,a4paper]{article}

\usepackage{epsfig}
\usepackage{amsmath,amsfonts,amssymb}
\usepackage{t1enc}
\usepackage{cite}

\providecommand{\openone}{\leavevmode\hbox{\small1\kern-3.8pt\normalsize1}}

\newcommand{\RE}{\text{Re}}

\newcommand{\Gp}{\Gamma_R}
\newcommand{\Gm}{\Gamma_L}
\newcommand{\Gz}{\Gamma_0}
\newcommand{\Gpm}{\Gamma_{R,L}}
\newcommand{\Fi}{F_i}
\newcommand{\Fpm}{F_{R,L}}
\newcommand{\fp}{F_R}
\newcommand{\fm}{F_L}
\newcommand{\fz}{F_0}
\newcommand{\rhp}{\rho_R}
\newcommand{\rhm}{\rho_L}
\newcommand{\rhpm}{\rho_{R,L}}

\newcommand{\thlw}{\theta_{\ell}^*}
\newcommand{\thlb}{\theta_{\ell b}}
\newcommand{\hk}{\alpha}

\newcommand{\afb}{A_\mathrm{FB}}
\newcommand{\Ap}{A_+}
\newcommand{\Am}{A_-}

\newcommand{\all}{A_{\ell \ell'}}
\newcommand{\anl}{A_{\nu \ell'}}
\newcommand{\alj}{A_{\ell j}}
\newcommand{\anj}{A_{\nu j}}
\newcommand{\alb}{A_{lb}}
\newcommand{\abb}{A_{bb}}
\newcommand{\allt}{\tilde A_{\ell \ell'}}

\newcommand{\aljt}{\tilde A_{\ell j}}

\newcommand{\vl}{V_L}
\newcommand{\vr}{V_R}
\newcommand{\gl}{g_L}
\newcommand{\gr}{g_R}
\newcommand{\qb}{|\vec q\,|}

\newcommand{\el}{E_\ell}
\newcommand{\emax}{E_\text{max}}
\newcommand{\emin}{E_\text{min}}
\newcommand{\lqcd}{\Lambda_\mathrm{QCD}}

\begin{document}

\begin{center}
\begin{Large}
{\bf Probing anomalous $\boldsymbol{Wtb}$ couplings in top pair decays}
\end{Large}

\vspace{0.5cm}
J. A. Aguilar--Saavedra$^a$, J. Carvalho$^b$, N. Castro$^b$, A. Onofre$^{b,c}$,
F. Veloso$^b$  \\[0.2cm] 
{\it $^a$ Departamento de Física Teórica y del Cosmos and CAFPE, \\
Universidad de Granada, E-18071 Granada, Spain} \\[0.1cm]
{\it $^b$ LIP - Departamento de Física, \\
Universidade de Coimbra, 3004-516 Coimbra, Portugal} \\[0.1cm]
{\it $^c$ UCP, Rua Dr. Mendes Pinheiro 24, 3080 Figueira da Foz, Portugal} \\
\end{center}

\begin{abstract}
We investigate several quantities, defined in the decays of top quark pairs,
which can be used to explore non-standard $Wtb$ interactions.
Two new angular asymmetries are introduced in the leptonic decay of top 
(anti)quarks. Both are
very sensitive to
anomalous $Wtb$ couplings, and their measurement allows for a precise
determination of the $W$ helicity fractions. We also examine other
angular and energy asymmetries, the $W$ helicity fractions and their ratios, as
well as spin correlation asymmetries, analysing their dependence on anomalous
$Wtb$ couplings and identifing the quantities which are most sensitive to them.
It is explicitly shown that spin correlation asymmetries are less sensitive
to new interactions in the decay of the top quark; therefore, when combined
with the measurement of other observables, they can be used to determine the
$t \bar t$ spin correlation even in the presence of anomalous $Wtb$ couplings.
We finally discuss some asymmetries which can be used to test CP
violation in $t \bar t$ production and complex phases in the effective $Wtb$
vertex.
\end{abstract}

\section{Introduction}

Precision studies have been in the past a powerful tool to explore new physics
at scales not kinematically accessible. With the operation of the Large Hadron
Collider (LHC), top physics will enter into the era of precise measurements
\cite{topquark}. Due to its large mass, close to the electroweak scale, the top
quark is believed to offer a unique window to physics beyond the Standard Model
(SM). New interactions at higher energies may manifest themselves in the form of
effective couplings of the SM fermions, especially for the top quark,
much heavier than the rest. In this work
we concentrate ourselves on the $Wtb$ vertex. Within the SM this coupling
is purely left-handed, and its size is given by the Cabibbo-Kobayashi-Maskawa
matrix element $V_{tb}$, which can be measured in single top production
\cite{single1,single2,single3}.
In new physics models, departures from the SM expectation 
$V_{tb} \simeq 1$ are possible \cite{largo,jose}, as well as new radiative
contributions to the $Wtb$ vertex \cite{susy,tech}. These corrections
can be parameterised with the effective operator formalism.
The most general $Wtb$ vertex containing terms up to dimension
five can be written as
\begin{eqnarray}
\mathcal{L} & = & - \frac{g}{\sqrt 2} \bar b \, \gamma^{\mu} \left( \vl
P_L + \vr P_R
\right) t\; W_\mu^- \nonumber \\
& & - \frac{g}{\sqrt 2} \bar b \, \frac{i \sigma^{\mu \nu} q_\nu}{M_W}
\left( \gl P_L + \gr P_R \right) t\; W_\mu^- + \mathrm{h.c.} \,,
\label{ec:lagr}
\end{eqnarray}
with $q=p_t-p_b$ (we follow the conventions of Ref.~\cite{prd} with slight
simplifications in the notation).
If CP is conserved in the decay, the couplings can be taken to be
real.\footnote{A general $Wtb$ vertex also contains terms proportional to
$(p_t+p_b)^\mu$, $q^\mu$ and $\sigma^{\mu \nu} (p_t+p_b)_\nu$. Since $b$ quarks
are on shell, the $W$ bosons decay to light particles (whose masses can be
neglected) and the top quarks can be approximately assumed on-shell, these
extra operators can be rewritten in terms of the ones in Eq.~(\ref{ec:lagr})
using Gordon identities.}
Within the SM, $\vl \equiv V_{tb} \simeq 1$ and $\vr$, $\gl$, $\gr$ vanish at
the tree level, while nonzero values are generated at one loop
level \cite{korner}. Additional contributions to $\vr$, $\gl$, $\gr$ are
possible in SM extensions, without spoiling the agreement with low-energy
measurements.
The size of a $\vr$ term  is constrained by the measured rate of
$\mathrm{Br}(b \to s \gamma) = (3.3 \pm 0.4) \times 10^{-4}$ \cite{PDB}.
A right-handed coupling $|\vr| \gtrsim 0.04$ would in principle give a too
large
contribution to this decay \cite{vtbrbound} which, however, might be (partially)
cancelled with other new physics contributions. Hence, the bound $|\vr|
\leq 0.04$ is model
dependent and does not substitute a direct measurement of this coupling. Similar
arguments
applied to the $\sigma^{\mu \nu}$ terms do not set relevant constraints on
$\gr$, because its contribution is suppressed by the ratio $q_\nu/M_W$ for
small $q_\nu$. 

Top production and decay processes at LHC allow us to probe the $Wtb$ vertex
\cite{single1,single3,prd,espriu,larios}.
Top pair production takes place through QCD interactions without involving
a $Wtb$ coupling. Additionally, it is likely that the top quark almost
exclusively decays in the channel $t \to W^+ b$.
Therefore, its cross section for production and decay $gg,q\bar q \to t \bar t
\to W^+ b W^- \bar b$ is insensitive to the size and structure of the $Wtb$
vertex. However, the angular distributions of (anti)top decay products give
information about its structure, and can then be used to trace
non-standard couplings. Angular distributions relating
top and antitop decay products probe not only the $Wtb$ interactions but also
the spin correlations among the two quarks produced, and thus may be influenced
by new production mechanisms as well.
On the other hand, single top production is sensitive to
both the size and structure of the $Wtb$ vertex, involved in the production
and the decay of the top quark \cite{single3,espriu,larios}.

In this paper we explore the sensitivity of several quantities, like angular and
energy asymmetries, helicity fractions and ratios, to new non-standard
$Wtb$ interactions. Although these observables are theoretically related,
the experimental determination is more precise for some of them than for others.
In particular, the experimental precision is dominated by systematics already
for a luminosity of 10 fb$^{-1}$, and a good choice of observables can improve
significantly the limits on anomalous $Wtb$ interactions.
Our analysis here is kept at a purely theoretical level,
identifying the quantities which are {\em a priori} more sensitive to anomalous
couplings, and estimating the precision in their experimental measurement from
a detailed simulation, which has been presented elsewhere
\cite{nuestro2,nuestro}.

\section{$\boldsymbol{W}$ helicity fractions and ratios}
\label{sec:2}

\begin{figure}[htb]
\begin{center}
\epsfig{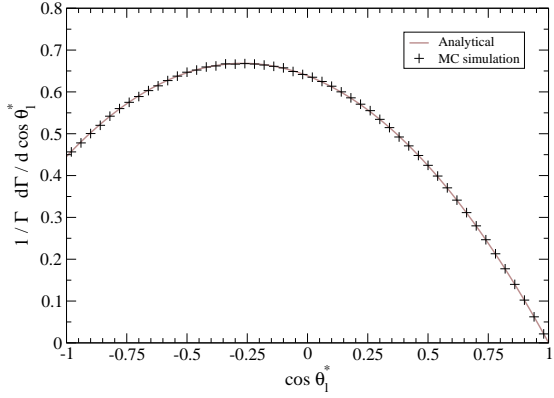}
\caption{Differential distribution in Eq.~(\ref{ec:dist}) within the SM,
calculated analytically and with a Monte Carlo simulation.}
\label{fig:dist-cos}
\end{center}
\end{figure}

The polarisation of the $W$ bosons emitted in the top decay is sensitive
to non-standard couplings \cite{kane}. The $W$ bosons can be
produced with positive (right-handed), negative (left-handed) or zero helicity,
with corresponding partial widths $\Gp$, $\Gm$, $\Gz$, being
$\Gamma \equiv \Gamma(t \to W^+ b) = \Gp + \Gm + \Gz$. The $\Gp$
component vanishes
in the $m_b = 0$ limit because the $b$ quarks produced in top decays have
left-handed chirality, and for vanishing $m_b$ the helicity and the chirality
states coincide. 
The three partial widths can be calculated for
a general $Wtb$ vertex as parameterised in Eq.~(\ref{ec:lagr}), yielding
\begin{eqnarray}
\Gz & = & \frac{g^2 \qb}{32 \pi} \left\{ \frac{m_t^2}{M_W^2} 
\left[ |\vl|^2 + |\vr|^2 \right] \left(1 - x_W^2 - 2 x_b^2 - x_W^2 x_b^2 
+ x_b^4 \right) - 4 x_b \, \RE \, \vl \vr^* \right. \notag \\
& & + \left[ |\gl|^2 + |\gr|^2 \right] \left(1 - x_W^2 + x_b^2 \right) 
- 4 x_b \, \RE \, \gl \gr^* \notag \\
& & - 2 \frac{m_t}{M_W} \RE \, \left[\vl \gr^* + \vr \gl^* \right]
\left(1 - x_W^2 - x_b^2 \right) \notag \\
& & \left. + 2 \frac{m_t}{M_W} x_b \,\RE \, \left[\vl \gl^* + \vr \gr^* \right]
\left(1 +x_W^2 - x_b^2 \right) \right\} \,, \notag \\
\Gpm & = & \frac{g^2 \qb}{32 \pi} \left\{
\left[ |\vl|^2 + |\vr|^2 \right] \left(1 - x_W^2 + x_b^2 \right) 
  - 4 x_b \, \RE \, \vl \vr^* \right. \notag \\
& & + \frac{m_t^2}{M_W^2} \left[ |\gl|^2 + |\gr|^2 \right] \left(1 - x_W^2 
- 2 x_b^2 - x_W^2 x_b^2 + x_b^4 \right) - 4 x_b \, \RE \, \gl \gr^* \notag \\
& & - 2 \frac{m_t}{M_W} \RE \, \left[\vl \gr^* + \vr \gl^* \right]
\left(1 - x_W^2 - x_b^2 \right) \notag \\
& & \left. + 2 \frac{m_t}{M_W} x_b \,\RE \, \left[\vl \gl^* + \vr \gr^* \right]
\left(1 +x_W^2 - x_b^2 \right) \right\}  \notag \\
& & \pm \frac{g^2}{64 \pi} \frac{m_t^3}{M_W^2} \left \{
 - x_W^2 \left[ |\vl|^2 - |\vr|^2 \right] 
 + \left[ |\gl|^2 - |\gr|^2 \right] \left(1 - x_b^2 \right) \right. 
  \notag \\
& & \left. + 2 x_W \, \RE \, \left[\vl \gr^* - \vr \gl^* \right] 
 + 2 x_W x_b \, \RE \, \left[\vl \gl^* - \vr \gr^* \right] \right\} \notag \\
& & \times \left( 1 - 2 x_W^2 - 2 x_b^2 + x_W^4 - 2 x_W^2 x_b^2 + x_b^4 \right)
\,,
\label{ec:gammas}
\end{eqnarray}
being $x_W = M_W/m_t$, $x_b = m_b/m_t$ and
\begin{equation}
\qb = \frac{1}{2 m_t} (m_t^4 + M_W^4 + m_b^4 - 2 m_t^2 M_W^2 
- 2 m_t^2 m_b^2 - 2 M_W^2 m_b^2)^{1/2}
\label{ec:qb}
\end{equation}
the modulus of the $W$ boson three-momentum in the top quark rest frame.
The total top width is
\begin{eqnarray}
\Gamma & = & \frac{g^2 \qb}{32 \pi} \frac{m_t^2}{M_W^2}
\left\{ \left[ |\vl|^2 + |\vr|^2 \right] \left(1 + x_W^2 - 2 x_b^2 -2 x_W^4
+ x_W^2 x_b^2 + x_b^4 \right)  \right.
 \notag \\
& & - 12 x_W^2 x_b \, \RE \, \vl \vr^* + 2 \left[ |\gl|^2 + |\gr|^2 \right]
\left(1 - \frac{x_W^2}{2}  - 2 x_b^2 - \frac{x_W^4}{2} - \frac{x_W^2 x_b^2}{2}
+ x_b^4 \right)   \notag \\
& & - 12 x_W^2 x_b \, \RE \, \gl \gr^* 
- 6 x_W \RE \, \left[\vl \gr^* + \vr \gl^* \right]
\left(1 - x_W^2 - x_b^2 \right) \notag \\
& & \left. + 6 x_W x_b \,\RE \, \left[\vl \gl^* + \vr \gr^* \right]
\left(1 +x_W^2 - x_b^2 \right) \right\} \,.
\end{eqnarray}
The different polarisation states of the $W$
boson are reflected in the angular distribution of its decay products.
Let us denote by $\thlw$ the angle between the charged lepton
three-momentum in the $W$ rest frame and the $W$ momentum in the $t$ rest frame.
The normalised differential decay rate for unpolarised top quarks can be
written as
\begin{equation}
\frac{1}{\Gamma} \frac{d \Gamma}{d \cos \thlw} = \frac{3}{8}
(1 + \cos \thlw)^2 \, \fp + \frac{3}{8} (1-\cos \thlw)^2 \, \fm
+ \frac{3}{4} \sin^2 \thlw \, \fz \,,
\label{ec:dist}
\end{equation}
with $\Fi \equiv \Gamma_i/\Gamma$ the helicity fractions. The three terms
correspond to the three
helicity states, and the interference terms vanish \cite{dalitz}. 
At the tree level, $\fz = 0.703$, $\fm = 0.297$,
$\fp = 3.6 \times 10^{-4}$, for $m_t = 175$ GeV, $M_W = 80.39$ GeV, $m_b = 4.8$
GeV. The resulting distribution is shown in
Fig.~\ref{fig:dist-cos}, calculated from the analytical expressions
in Eqs.~(\ref{ec:gammas})--(\ref{ec:dist}) and also with a Monte Carlo
simulation. The latter is performed using our own $t \bar t$ generator,
which uses the full resonant matrix element for
$gg, q \bar q \to t \bar t \to W^+ b W^- \bar b \to f_1 \bar f_1' b \bar f_2
f_2' \bar b$, and hence takes into account the top and $W$ widths, as well as
their polarisations. Anomalous couplings in the decay may also included in the
event generation.
We observe that finite width corrections have a negligible
influence in the distribution, and hence Eqs.~(\ref{ec:gammas})--(\ref{ec:dist}) 
can be used to make precise predictions for the distributions.

The good agreement between the analytical calculation (with the top quark and
$W$ boson on their mass shell) and the numerical one can be
explained substituting $m_t \to m_t (1+ \xi_t \, \Gamma_t/m_t)$,
$M_W \to M_W (1 + \xi_W \, \Gamma_W/M_W)$, with $\xi_t$, $\xi_W$ of order unity,
in the expressions of the helicity
fractions (here we introduce subscripts to distinguish the top quark and $W$ 
boson widths). We obtain 
\begin{align}
\fz & = 0.703 \left[ 1 + 0.597 \left( \xi_t \frac{\Gamma_t}{m_t} \right) 
 - 0.595 \left( \xi_W \frac{\Gamma_W}{M_W} \right)  
 - 0.545 \left( \xi_t \frac{\Gamma_t}{m_t} \right)^2 
  \right. \notag \\
& \left.  + 0.055 \left( \xi_W \frac{\Gamma_W}{M_W} \right)^2
+ 0.487 \left( \xi_t \frac{\Gamma_t}{m_t} \right)
  \left( \xi_W \frac{\Gamma_W}{M_W} \right)
 + \dots \right] \,, \notag \\
\fp & = 3.6 \times 10^{-4} \left[ 1 
 - 4.48 \left( \xi_t \frac{\Gamma_t}{m_t} \right) 
 + 2.48 \left( \xi_W \frac{\Gamma_W}{M_W} \right)  
 + 13.22 \left( \xi_t \frac{\Gamma_t}{m_t} \right)^2 
  \right. \notag \\
& \left.  + 2.77 \left( \xi_W \frac{\Gamma_W}{M_W} \right)^2
- 12.98 \left( \xi_t \frac{\Gamma_t}{m_t} \right)
  \left( \xi_W \frac{\Gamma_W}{M_W} \right)
 + \dots \right] \,, \notag \\
\fm & = 0.297 \left[ 1 - 1.407 \left( \xi_t \frac{\Gamma_t}{m_t} \right) 
 + 1.405 \left( \xi_W \frac{\Gamma_W}{M_W} \right)  
 + 1.274 \left( \xi_t \frac{\Gamma_t}{m_t} \right)^2 
  \right. \notag \\
& \left.  - 0.134 \left( \xi_W \frac{\Gamma_W}{M_W} \right)^2
- 1.137 \left( \xi_t \frac{\Gamma_t}{m_t} \right)
  \left( \xi_W \frac{\Gamma_W}{M_W} \right)
 + \dots \right] \,.
\end{align}
Linear terms have no effect when integrated with symmetric Breit-Wigner
distributions, and the quadratic terms are very small.

\begin{figure}[t*]
\begin{center}
\begin{tabular}{ccc}
\epsfig{file=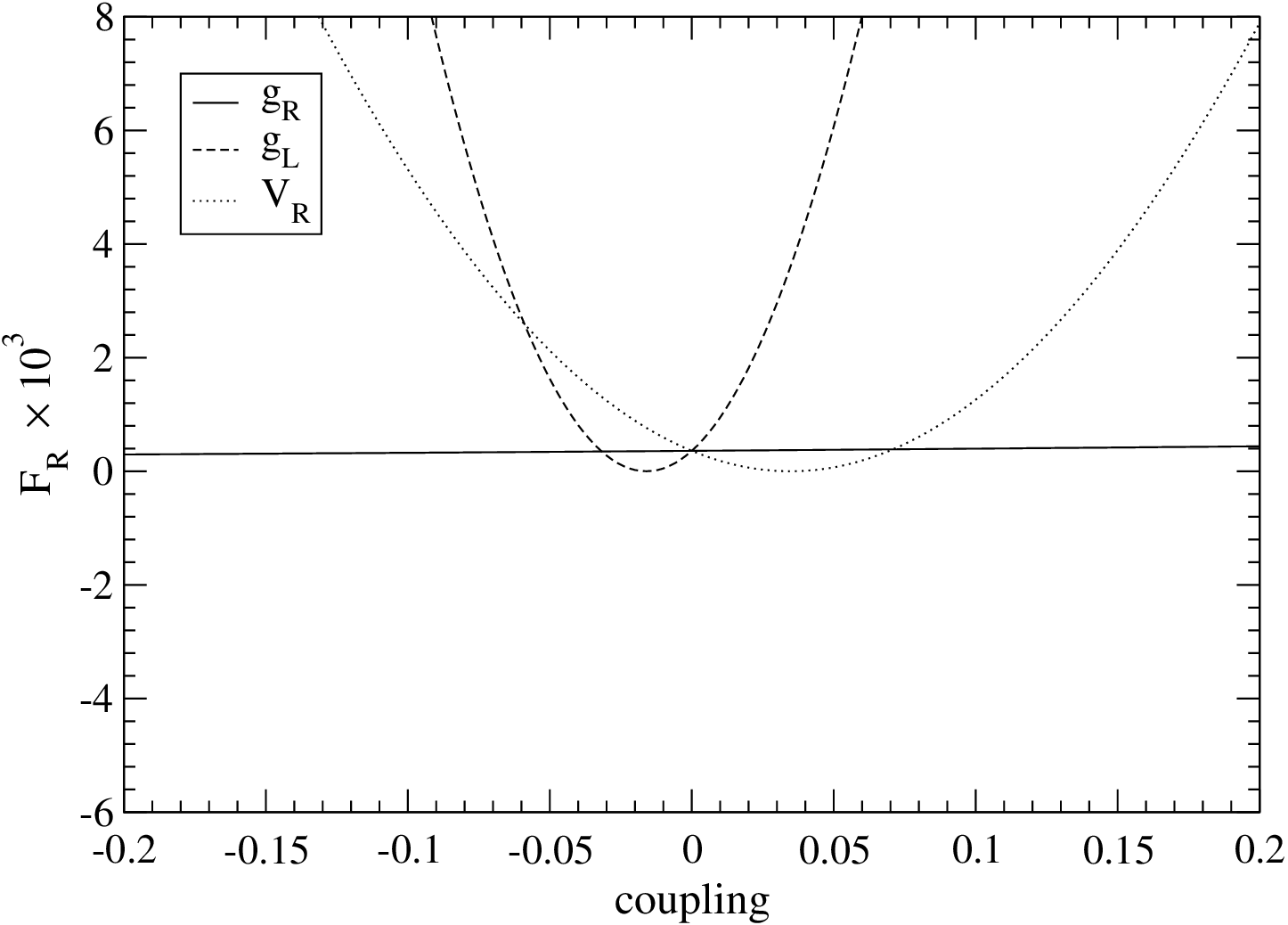,height=4.8cm,clip=} & ~ &
\epsfig{file=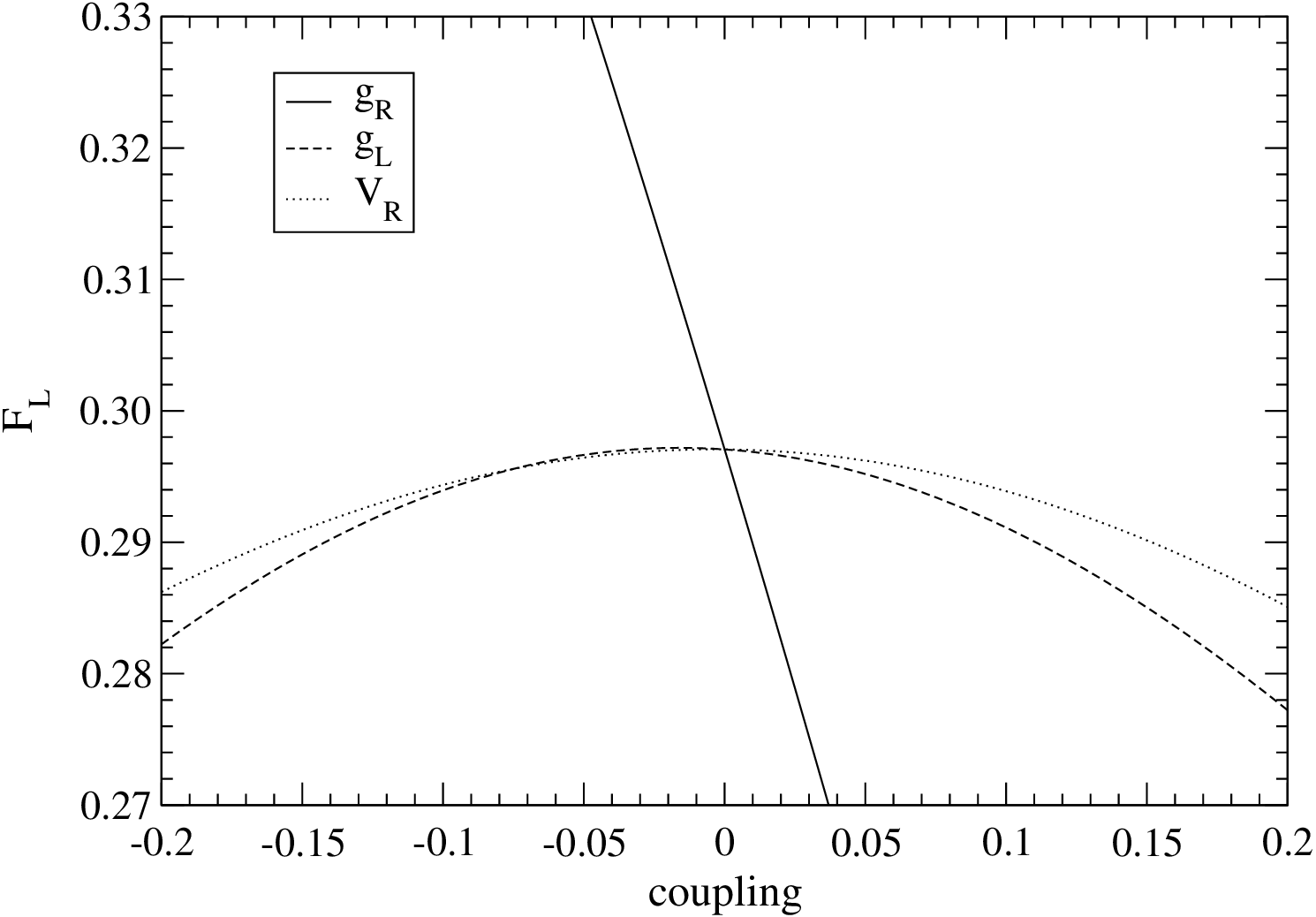,height=4.8cm,clip=} \\[0.5cm]
\multicolumn{3}{c}{\epsfig{file=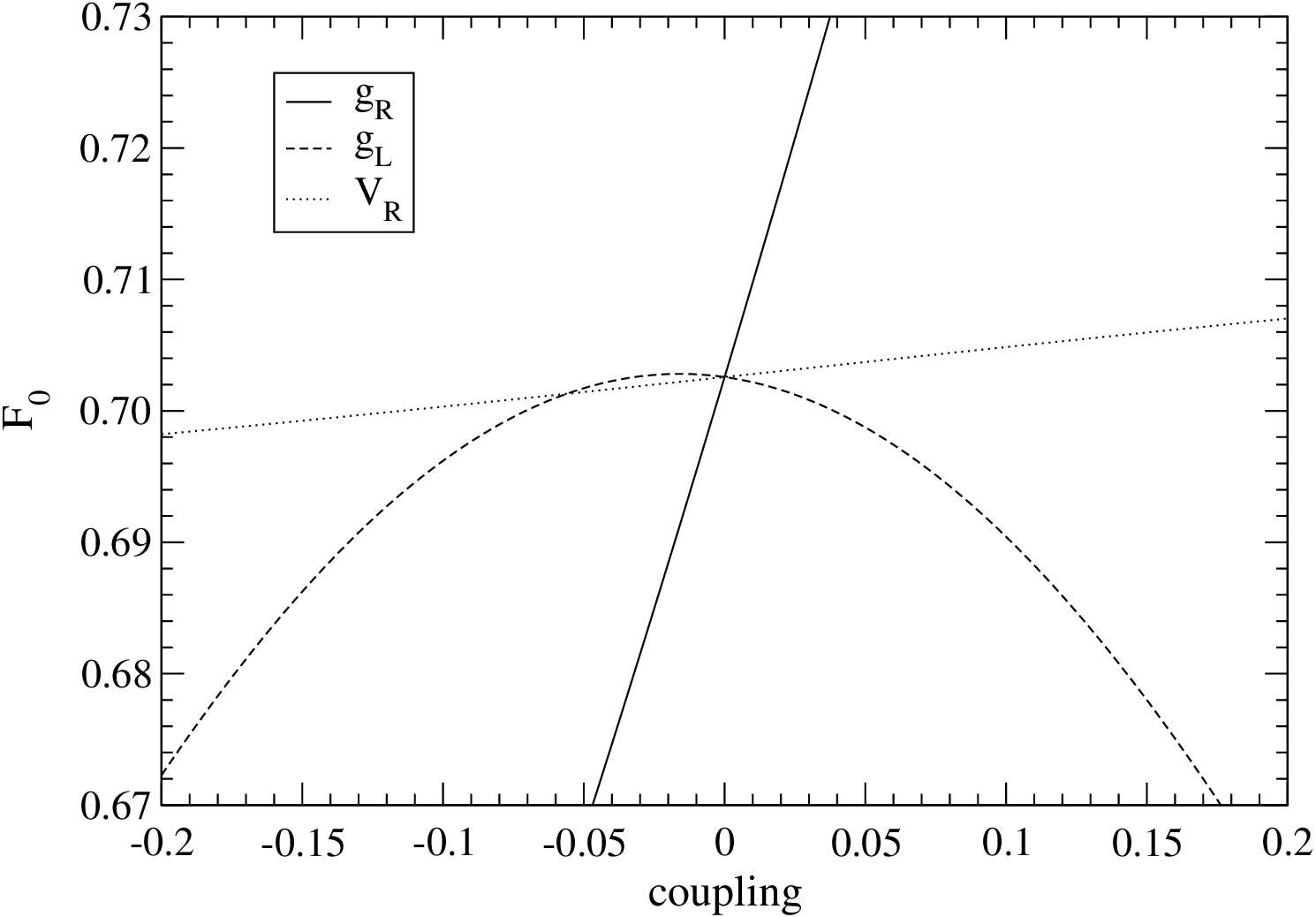,height=4.8cm,clip=}}
\end{tabular}
\caption{Dependence of the helicity fractions $\Fi = \Gamma_i/\Gamma$
on the anomalous couplings in Eq.~(\ref{ec:lagr}), in the CP-conserving case.}
\label{fig:widths}
\end{center}
\end{figure}

In the presence of anomalous couplings, the helicity fractions $\Fi$ are
modified with respect to their SM values quoted above. Their variation is
plotted in Fig.~\ref{fig:widths}, considering that only one coupling is
different from zero at a time and restricting ourselves to the CP-conserving
case of real $\vr$, $\gr$ and $\gl$. We observe that $\fm$ and $\fz$ are much
more sensitive to $\gr$ than to $\gl$ and $\vr$. This is due to the interference
term $\vl \gr^*$ , which is not suppressed by the bottom quark mass as for the
$\gl$ and $\vr$ couplings.
This linear term dominates over the quadratic one and makes the helicity
fractions (and related quantities) very sensitive to $\gr$. We also remark
that the phases of anomalous couplings
influence the helicity fractions through the interference terms
which depend on the real part of $\vr$,
$\gl$ and $\gr$ (we have taken $\vl$ real, and normalised to unity). Thus,
the effect of complex phases is specially relevant for $\gr$, where the
interference term dominates. In any case, the maximum and
minimum deviations on the helicity fractions are found for real, positive and
negative (not necessarily in this order) values of $\vr$, $\gr$ and $\gl$.
The possibility of complex couplings is examined with more detail in
section~\ref{sec:a}.

The helicity fractions can be experimentally extracted from a fit to the
$\cos \thlw$ distribution using Eq.~(\ref{ec:dist}). In order to estimate the
limits on anomalous couplings that can be set from their measurement, we assume
that the central values obtained correspond to the SM prediction, and take
their errors from Refs.~\cite{marsella,nuestro2,nuestro}, giving
$\fz \simeq 0.703 \pm 0.016$, $\fp \simeq 3.6 \times 10^{-4} \pm 0.0045$,
$\fm \simeq 0.297 \pm 0.016$.
For these values, it is found
that $\fz$ and $\fm$ have a similar sensitivity
to $\gr$, while the dependence of $\fp$ on this coupling is smaller. On the
other hand, the measurement of $\fp$ sets the strongest constraint on $\vr$ and
$\gl$. The resulting bounds are summarised in the first column of
Table~\ref{tab:limits1}. These and the rest of limits throughout this paper
have been obtained with a Monte Carlo method, as
described in appendix \ref{sec:c}.

\begin{table}[htb]
\begin{center}
\begin{tabular}{ccc}
& $\Fi$ & $\rho_i$ \\
\hline
$\vr$ & $[-0.062 , 0.13]$ & $[-0.029 , 0.099]$ \\
$\gl$ & $[-0.060 , 0.028]$ & $[-0.046 , 0.013]$ \\
$\gr$ & $[-0.023 , 0.021]$ & $[-0.025 , 0.026]$ \\
\hline
\end{tabular}
\caption{$1\sigma$ bounds of anomalous couplings obtained from the
measurement of helicity fractions $\Fi$ and ratios $\rho_i$.}
\label{tab:limits1}
\end{center}
\end{table}

\begin{figure}[t*]
\begin{center}
\begin{tabular}{ccc}
\epsfig{file=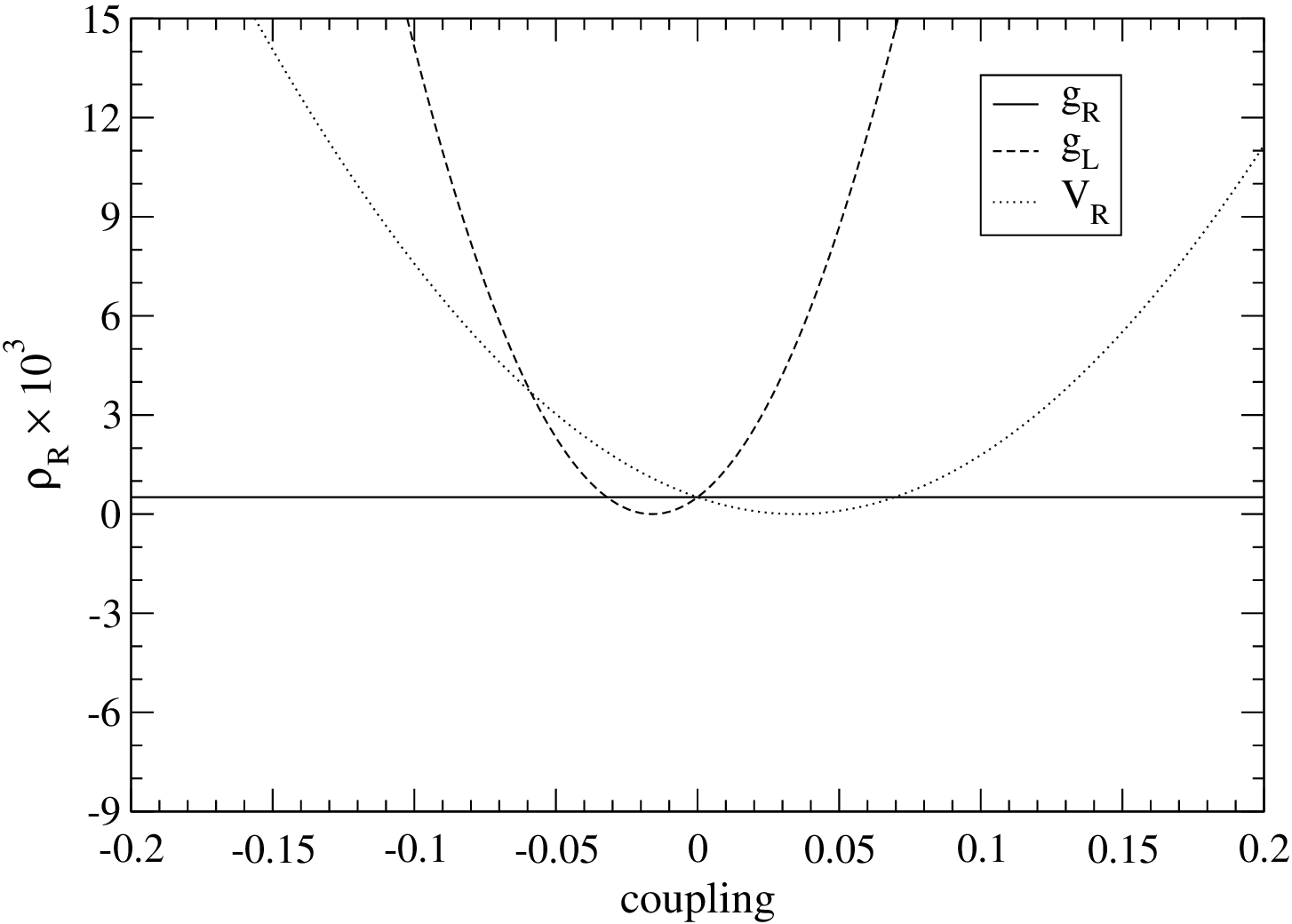,height=4.8cm,clip=} & ~ &
\epsfig{file=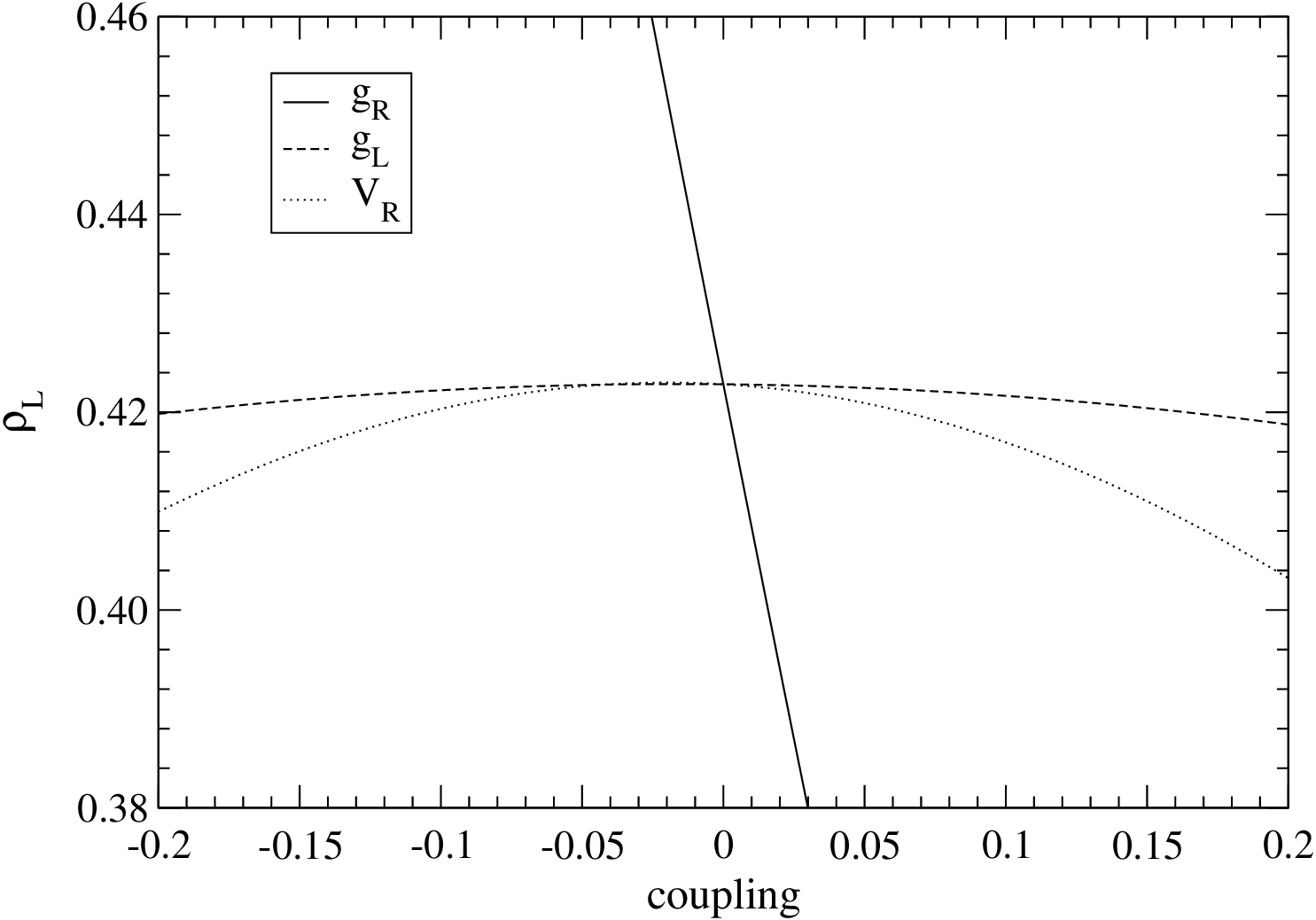,height=4.8cm,clip=} 
\end{tabular}
\caption{Dependence of the helicity ratios $\rhpm = \Gpm/\Gz$ on the
anomalous couplings in Eq.~(\ref{ec:lagr}), in the CP-conserving case.}
\label{fig:rhos}
\end{center}
\end{figure}

The sensitivity achieved for non-standard couplings  may be greater if we
consider instead the helicity ratios
$\rhpm \equiv \Gpm / \Gz = \Fpm / \fz$,
shown in Fig.~\ref{fig:rhos}.\footnote{We note that, for a better comparison
among them and with other observables, the scale of the $y$ axis in each plot 
is chosen so
that the range approximately corresponds to  two standard deviations (with the
expected LHC precision) around the theoretical SM value.}
These ratios can be
directly measured with a fit to the $\cos \thlw$ distribution as well.
From the expected precision in their determination in
Ref.~\cite{nuestro2,nuestro}, and assuming that the central values correspond
to the SM prediction, we have
$\rhp \simeq 0.0005 \pm 0.0026$, $\rhm \simeq 0.423 \pm 0.036$.
From these values, the limits given in the second column of
Table~\ref{tab:limits1} can be obtained, with an important improvement for
$\vr$ and $\gl$. As it has been remarked in the introduction, the reason for the
improvement is that systematic errors, which dominate the precision of the
measurements (see Refs.~\cite{nuestro2,nuestro} for details), are much smaller
for helicity ratios than for helicity fractions.

To conclude this section, we would like to stress the importance of keeping
the bottom quark mass in the calculations. Within the SM the $m_b$
correction to the helicity fractions is small, of order $x_b^2 = 7.5 \times
10^{-4}$, as it can be seen in Eqs.~(\ref{ec:gammas}). However, as it can also
be observed, the interference terms involving $\gl$ or $\vr$ couplings with
$\vl$ are proportional to $x_b = 0.027$, and are of
similar magnitude as the quadratic terms. The effect of
including $m_b$ in the computations is illustrated with more detail in
appendix \ref{sec:b}. Nevertheless, we note here that if $m_b$ is neglected the
resulting confidence intervals on $\vr$, $\gl$ are symmetric.
The asymmetry between positive and negative couplings
seen in Table~\ref{tab:limits1} reflects the importance of the $m_b$ correction.
It should also be noted that the $m_b$ dependence of the limits leads to a small
systematic uncertainty, due to the uncertainty in $m_b$. This is examined in
detail in appendix \ref{sec:b}.

\section{Angular asymmetries}
\label{sec:3}

A simple and efficient method to extract information about the $Wtb$ vertex is
through angular
asymmetries involving the angle $\thlw$ between the charged lepton momentum (in
the $W$ boson rest frame) and the $W^+$ boson momentum (in the top quark rest
frame). Alternatively, one may consider the angle $\thlb$
between the charged lepton and $b$ quark momenta in the $W$ rest frame.
Both approaches are equivalent since
these two angles are related by $\thlw + \thlb = \pi$. (The determination of
$\thlb$, however, is simpler, because both momenta are measured in the same
reference frame without any ambiguity in the boosts.)
For any fixed $z$ in the interval $[-1,1]$, one can define an asymmetry
\begin{equation}
A_z = \frac{N(\cos \thlw > z) - N(\cos \thlw < z)}{N(\cos \thlw > z) +
N(\cos \thlw < z)} \,.
\end{equation}
The most obvious choice is $z=0$, giving the forward-backward (FB) asymmetry
$\afb$ \cite{lampe,prd}.\footnote{Notice the difference in sign with respect
to the definitions in Refs.~\cite{lampe,prd}, where $\thlb$ is used.}
It is analogous to the FB asymmetries at LEP, which together with the ratios
$R_b$, $R_c$ allow us to extract the couplings of the $c$ and $b$ quarks to the
$Z$ boson. The FB asymmetry is related to the $W$ helicity fractions by
\begin{equation}
\afb = \frac{3}{4} [\fp - \fm] \,.
\end{equation}
The measurement of this asymmetry alone is not enough to fully
reconstruct the $\cos \thlw$ distribution. One can then think about other
asymmetries for different values of $z$. The determination of $\Fi$ is easier
if we construct asymmetries involving only $\fp$ and $\fz$, or $\fm$ and $\fz$.
This is achieved choosing $z = \mp (2^{2/3}-1)$. Defining for convenience
$\beta = 2^{1/3}-1$, we have
\begin{eqnarray}
z = -(2^{2/3}-1) & \rightarrow & A_t = \Ap = 3 \beta [\fz+(1+\beta) \fp] \,,
\notag \\
z = (2^{2/3}-1) & \rightarrow & A_t = \Am = -3 \beta [\fz+(1+\beta) \fm] \,.
\end{eqnarray}
From both asymmetries and using $\fp+\fm+\fz = 1$, we obtain
\begin{eqnarray}
\fp & = & \frac{1}{1-\beta} + \frac{\Am - \beta \Ap}{3 \beta(1-\beta^2)} \,,
 \notag \\
\fm & = & \frac{1}{1-\beta} - \frac{\Ap - \beta \Am}{3 \beta(1-\beta^2)} \,,
 \notag \\
\fz & = & - \frac{1+\beta}{1-\beta} + \frac{\Ap - \Am}{3 \beta (1-\beta)} \,.
\end{eqnarray}

\begin{figure}[t*]
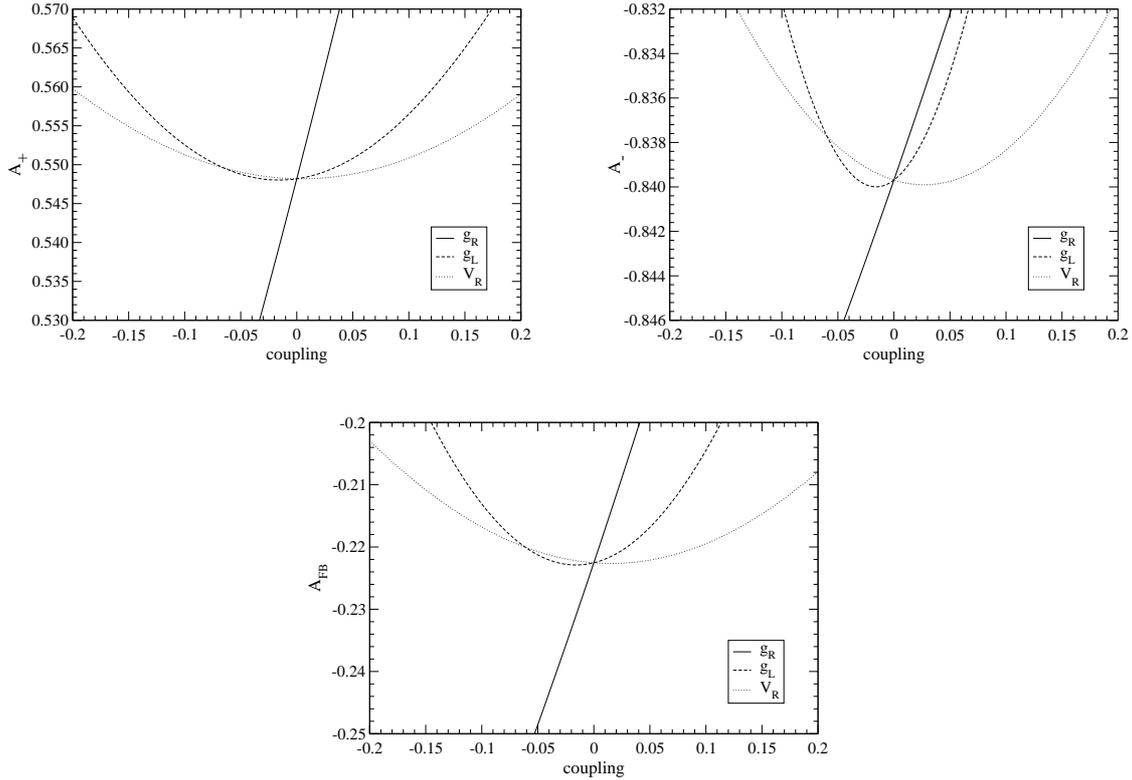

\begin{center}
\begin{tabular}{ccc}
\epsfig{file=Figs/ap.eps,height=4.8cm,clip=} & ~ &
\epsfig{file=Figs/am.eps,height=4.8cm,clip=} \\[0.5cm]
\multicolumn{3}{c}{\epsfig{file=Figs/afb.eps,height=4.8cm,clip=}}
\end{tabular}
\caption{Dependence of the asymmetries $\Ap$, $\Am$ and $\afb$ on the couplings
$\gl$, $\gl$ and $\vr$, for the CP-conserving case.}
\label{fig:asim}
\end{center}
\end{figure}

The three asymmetries $\afb$, $\Ap$, $\Am$ are quite sensitive to anomalous
$Wtb$ interactions. Their SM values are $\afb = -0.2225$, $\Ap = 0.5482$, $\Am =
-0.8397$, and their dependence on the non-standard couplings is shown in
Fig.~\ref{fig:asim}. Taking the expected precision in their measurement
from Refs.~\cite{nuestro2,nuestro} and assuming as central values the SM
predictions, we obtain
$\afb \simeq -0.223 \pm 0.013$, $\Ap \simeq 0.548 \pm 0.010$,
$\Am \simeq -0.8397 \pm 0.0033$. Using {\em e.g.} the latter two,
the helicity fractions can be determined as
\begin{eqnarray}
\fp & = &  0.0017 \pm 0.0071 \,, \notag \\
\fm & = & 0.2981 \pm 0.0167 \,, \notag \\
\fz & = & 0.7002 \pm 0.0184 \,.
\end{eqnarray}
The errors quoted take into account the correlation between the two
measurements, which is determined writing the asymmetries in terms of the
numbers of events in three bins: $[-1,-(2^{2/3}-1)]$,
$[-(2^{2/3}-1),(2^{2/3}-1)]$ and $[(2^{2/3}-1),1]$. We omit these details for
brevity. The values extracted in this way are less precise than if obtained from
a direct fit, but the method employed is much simpler too.
The eventual limits which would
be extracted from asymmetry measurements are collected in
Table~\ref{tab:limits2}.
$\Ap$ exhibits the strongest dependence on $\gr$ and, if measured as precisely
as it is expected, it would set the best limits on this coupling. On the other
hand, $\Am$ is the most sensitive to $\vr$ and $\gl$ and sets the strongest
bounds on them. The limits obtained from asymmetry measurements are
competitive with those obtained from a direct fit to the $\cos \thlw$
distribution.

\begin{table}[htb]
\begin{center}
\begin{tabular}{cccc}
& $\Ap$ & $\Am$ & $\afb$ \\
\hline
$\vr$ & $[-0.15 , 0.15]$   & $[-0.056 , 0.11]$  & $[-0.12 , 0.15]$ \\
$\gl$ & $[-0.12 , 0.082]$  & $[-0.057 , 0.026]$ & $[-0.092 , 0.062]$ \\
$\gr$ & $[-0.019 , 0.018]$ & $[-0.024 , 0.022]$ & $[-0.027 , 0.025]$ \\
\hline
\end{tabular}
\caption{$1\sigma$ bounds on anomalous couplings obtained from the
measurement of angular asymmetries.}
\label{tab:limits2}
\end{center}
\end{table}

\section{Energy distributions}
\label{sec:4}

\begin{figure}[htb]
\begin{center}
\epsfig{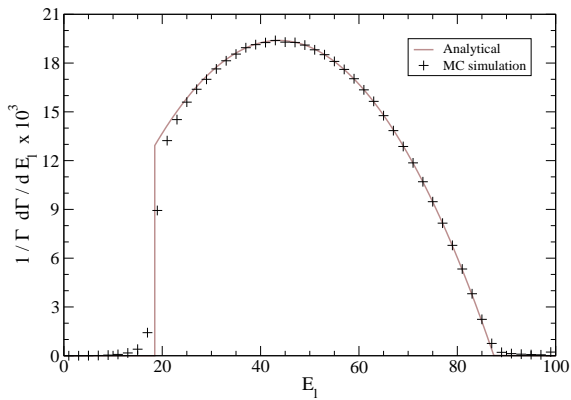}
\caption{Differential distribution in Eq.~(\ref{ec:dist2}) within the SM,
calculated analytically and with a Monte Carlo simulation.}
\label{fig:dist-el}
\end{center}
\end{figure}

The charged lepton energy in the $W$ rest frame is fixed by the kinematics of
the two-body decay $W \to \ell \nu$. Its energy in the top quark rest frame,
denoted from now on by $E_\ell$, is related to the former by a Lorentz boost,
and it is given by
\begin{equation}
\el = \frac{1}{2} (E_W + \qb \cos \thlw) \,,
\label{ec:el}
\end{equation}
with $\qb$, given in Eq.~(\ref{ec:qb}), the $W$ boson momentum in the top rest
frame and $E_W$ its energy.
Therefore, the angular distribution of the charged lepton in $W$ rest frame
determines its energy in the top rest frame. The maximum and minimum energies
are $\emax = (E_W + \qb)/2$, $\emin = (E_W - \qb)/2$.
The energy distribution is
obtained from Eqs.~(\ref{ec:dist}) and (\ref{ec:el}),
\begin{eqnarray}
\frac{1}{\Gamma} \frac{d \Gamma}{d \el} & = & \frac{1}{(\emax -
\emin)^3} \left[ 3 (\el - \emin)^2 \, \fp + 3 (\emax - \el)^2 \, \fm
\right. \notag \\[1mm]
& & \left. + 6 (\emax - \el) (\el - \emin) \, \fz \right] \,.
\label{ec:dist2}
\end{eqnarray}
The description of the top decay in terms of $\cos \thlw$ or $\el$ seems then
equivalent, up to a change of variables. Any asymmetry built using $\cos
\thlw$, defined around a fixed value $z$, can be translated into an equivalent
asymmetry involving $\el$, defined around a fixed energy
$E_z = (E_W + \qb \, z)/2$, namely
\begin{equation}
A_z = \frac{N(\el > E_z) - N(\el < E_z)}{N(\el > E_z) + N(\el < E_z)} \,.
\end{equation}
However,
in contrast to what was demonstrated in section~\ref{sec:2} for the angular
distributions, finite width corrections have a non-negligible influence on
$\el$. This can be seen in Fig.~\ref{fig:dist-el}, where we plot the energy
distribution calculated analytically for $t$ and $W$ on shell and from a Monte
Carlo calculation including finite width effects.
The values of the asymmetries $\Ap$, $\Am$ and $\afb$, calculated analytically
and numerically (the latter for the $\cos \thlw$ and $\el$ distributions) are
shown in Table~\ref{tab:asimcomp}. One can notice the larger influence of finite
width corrections for energy asymmetries.

\begin{table}[htb]
\begin{center}
\begin{tabular}{cccc}
& Analytical & Angular & Energy \\
\hline
$\Ap$  &  0.5482 &  0.5492 &  0.5529 \\
$\Am$  & -0.8397 & -0.8393 & -0.8339 \\
$\afb$ & -0.2225 & -0.2212 & -0.2166 \\
\hline
\end{tabular}
\caption{Values of the asymmetries $\Ap$, $\Am$, $\afb$ obtained from the
analytical expression (first column) and from the Monte Carlo simulation, the
latter from the measurement of the distributions of $\cos \thlw$ (second column)
and $E_l$ (third column).}
\label{tab:asimcomp}
\end{center}
\end{table}

The expected precision in energy asymmetries is worse than for the angular ones
(given in the previous section) as it might be expected:
$\afb \simeq -0.223 \pm 0.024$, $\Ap \simeq 0.548 \pm 0.013$,
$\Am \simeq -0.840 \pm 0.016$. Therefore, in principle their
study does not seem to bring any
improvement from the experimental side.

\section{Spin asymmetries}
\label{sec:5}

Additional angular asymmetries can be built involving the top spin. Top
quarks are produced unpolarised at the tree level in QCD interactions, and with
a very small $O(10^{-2})$ transverse polarisation at one loop. However, the $t$
and $\bar t$ spins are strongly correlated, what allows the construction of
angular asymmetries at the percent level. The top (anti)quark spins are not
directly observable, but influence the angular distribution of their decay
products.
For the decay $t \to W^+ b \to \ell^+ \nu b,q \bar q' b$, the angular
distributions of $X=\ell^+,\nu,q,\bar q',W^+,b$ (which are
called ``spin analysers'') in the top quark rest frame are given by
\begin{equation}
\frac{1}{\Gamma} \frac{d\Gamma}{d \cos \theta_X} = \frac{1}{2} (1+\hk_X \cos
\theta_X )
\label{ec:tdist}
\end{equation}
with $\theta_X$ the angle between the three-momentum of $X$ (in the $t$
rest frame) and the top spin direction.
The constants $\hk_X$ are called ``spin analysing power'' of $X$ and can range
between $-1$ and $1$. In the SM,
$\hk_{\ell^+} = \hk_{\bar q'} = 1$,
$\hk_\nu = \hk_q = -0.32$, $\hk_{W^+} = - \hk_b = 0.41$ at the tree level
\cite{jezabek} ($q$ and $q'$ are the up- and down-type quarks, respectively,
resulting from the $W$ decay). 
For the decay of a top antiquark the distributions are the same, with
$\hk_{\bar X} = - \hk_X$ as long as CP is conserved in the decay. 
One-loop corrections modify these values to
$\hk_{\ell^+} = 0.998$, $\hk_{\bar q'} = 0.93$,
$\hk_\nu = -0.33$, $\hk_q = -0.31$, $\hk_{W^+} = - \hk_b = 0.39$
\cite{hqcd,bern}.
We point out that in the presence of
non-vanishing $\vr$, $\gl$ or $\gr$ couplings the numerical values of the
constants $\hk_X$ are modified, but the functional form of Eq.~(\ref{ec:tdist})
is maintained. We have explicitly calculated them for a general CP-conserving
$Wtb$ vertex as written in Eq.~(\ref{ec:lagr}) within the narrow width
approximation. They can be written as $\alpha_X = a_X/a_0$, with
\begin{align}
a_0 & = 
 \left[ \vl^2 + \vr^2 \right] \left(1 + x_W^2 - 2 x_b^2 -2 x_W^4
+ x_W^2 x_b^2 + x_b^4 \right) - 12 x_W^2 x_b \, \vl \vr \notag \\
&  + 2 \left[ \gl^2 + \gr^2 \right]
\left(1 - \frac{x_W^2}{2}  - 2 x_b^2 - \frac{x_W^4}{2} - \frac{x_W^2 x_b^2}{2}
+ x_b^4 \right)   \notag \\
& - 12 x_W^2 x_b \, \gl \gr 
- 6 x_W \left[\vl \gr + \vr \gl \right]
\left(1 - x_W^2 - x_b^2 \right) \notag \\
& + 6 x_W x_b \, \left[\vl \gl + \vr \gr \right]
\left(1 +x_W^2 - x_b^2 \right)  \,, \notag \\
a_{\ell^+,\bar q'} & =
   \left[ \vl^2 - \vr^2 \right] \left(1 + x_W^2 - 2 x_b^2 -2 x_W^4
  + x_W^2 x_b^2 + x_b^4 \right) - 12 x_W^2 x_b \, \vl \vr \notag \\
& + 2 \left[ \gl^2 - \gr^2 \right]
\left(1 - \frac{x_W^2}{2}  - 2 x_b^2 - \frac{x_W^4}{2} - \frac{x_W^2 x_b^2}{2}
+ x_b^4 \right) \notag \\
& - 12 x_W^2 x_b \, \gl \gr 
- 6 x_W \left[\vl \gr + \vr \gl \right]
\left(1 - x_W^2 - x_b^2 \right) \notag \\
& + 6 x_W x_b \, \left[\vl \gl - \vr \gr \right]
\left(1 +x_W^2 - x_b^2 \right) + 12 x_W^2 (\vr^2 - \gr^2) \notag \\
& + \frac{m_t}{|\vec q|} \log \frac{E_W+|\vec q|}{E_W-|\vec q|}
\left[ -6 x_W^4 \vr^2 + 6 x_W^2 \gr^2 (1-x_b^2) +12 x_W^3 x_b \vr \gr
\right] \,, \notag \\
a_{\nu, q} & = \left[ \vl^2 - \vr^2 \right] \left(1 + x_W^2 - 2 x_b^2 -2 x_W^4
+ x_W^2 x_b^2 + x_b^4 \right) + 12 x_W^2 x_b \, \vl \vr \notag \\
& + 2 \left[ \gl^2 - \gr^2 \right]
\left(1 - \frac{x_W^2}{2}  - 2 x_b^2 - \frac{x_W^4}{2} - \frac{x_W^2 x_b^2}{2}
+ x_b^4 \right) \notag \\
& + 12 x_W^2 x_b \, \gl \gr 
+ 6 x_W \left[\vl \gr + \vr \gl \right]
\left(1 - x_W^2 - x_b^2 \right) \notag \\
& + 6 x_W x_b \, \left[\vl \gl - \vr \gr \right]
\left(1 +x_W^2 - x_b^2 \right) - 12 x_W^2 (\vl^2 - \gl^2) \notag \\
& + \frac{m_t}{|\vec q|} \log \frac{E_W+|\vec q|}{E_W-|\vec q|}
\left[ 6 x_W^4 \vl^2 - 6 x_W^2 \gl^2 (1-x_b^2) -12 x_W^3 x_b \vl \gl
\right] \,, \notag \\
a_b & = -2 \frac{|\vec q|}{m_t} \left\{ \left[ \vl^2 - \vr^2 \right] 
\left( 1-2 x_W^2 - x_b^2 \right)
+ 2 \left[ \gl^2 - \gr^2 \right] \left( 1 - \frac{x_W^2}{2} +x_b^2 \right)
\right. \notag \\
& \left. +2 x_W \left[ \vl \gr - \vr \gl \right]
+ 6 x_W x_b \left[ \vl \gl - \vr \gr \right] \right\}   \,,
\label{ec:kappas}
\end{align}
and $a_W = -a_b$.
We have checked that our expressions are compatible with the first-order
expansions in Refs.~\cite{grzad1,grzad2}.
Working in the helicity basis and neglecting small spin interference
effects, so that the cross section factorises into production times decay
factors, the double angular distribution of the decay products $X$ (from $t$)
and $\bar X'$ (from $\bar t$) can be written as \cite{stelzer}
\begin{equation}
\frac{1}{\sigma} \frac{d\sigma}{d \cos \theta_X \, d \cos \theta_{\bar X'}} =
\frac{1}{4} (1+C \, \hk_X \hk_{\bar X'} \cos \theta_X \cos \theta_{\bar X'}) \,.
\label{ec:dbldist}
\end{equation}
The angles $\theta_X$, $\theta_{\bar X'}$ are measured using as spin axis
the parent top (anti)quark momentum in the $t \bar t$ CM system. The factor
\begin{equation}
C \equiv \frac{\sigma(t_R \bar t_R) + \sigma(t_L \bar t_L) -
\sigma(t_R \bar t_L) - \sigma(t_L \bar t_R)}{\sigma(t_R \bar t_R) +
\sigma(t_L \bar t_L) + \sigma(t_R \bar t_L) + \sigma(t_L \bar t_R)}
\label{ec:C}
\end{equation}
is the relative number of like helicity minus opposite helicity $t \bar t$
pairs, and measures the spin correlation between the top quark and
antiquark.\footnote{Other conventions in the literature ({\em e.g.}
Refs.~\cite{mahlon,bern}) denote by $-C$ what in our case is the product 
$C \, \hk_X \hk_{\bar X'}$. We prefer to keep the notation in
Refs.~\cite{stelzer,topquark} and separate the contributions from the
production ($C$)
and the decay ($\hk_X$, $\hk_{\bar X'}$) since the former is sensitive to new
physics in the $t \bar t$ production process while the latter are sensitive to
non-standard $Wtb$ interactions. This decomposition is not possible if
non-factorisable radiative corrections to the production and decay process are
included. Anyway, these corrections are expected to be small.}
We note that due to P invariance of the QCD
interactions,
$\sigma(t_R \bar t_L) = \sigma(t_L \bar t_R)$, and by CP conservation
$\sigma(t_R \bar t_R) = \sigma(t_L \bar t_L)$. This is the reason why 
terms linear in $\cos \theta_X$, $\cos \theta_{\bar X'}$ are absent in
Eq.~(\ref{ec:dbldist}). In other words, terms linear in the cosines are present
only if the top quarks are produced with a net polarisation in the helicity
basis, what does not happen in pure QCD production. The actual value of $C$
depends to
some extent on the parton distribution functions (PDFs) used and the $Q^2$ scale
at which they are evaluated.
Using the CTEQ5L PDFs \cite{cteq} and $Q^2 =\hat s$ the partonic CM energy,
we find $C = 0.310$.
At the one loop level, $C = 0.326 \pm 0.012$ \cite{bern}.

Using the spin analysers $X$, $\bar X'$ for the respective decays of $t$, $\bar
t$, one can define the asymmetries
\begin{equation}
A_{X \bar X'} \equiv \frac{N(\cos \theta_X \cos \theta_{\bar X'} > 0) 
- N(\cos \theta_X \cos \theta_{\bar X'} < 0)}
{N(\cos \theta_X \cos \theta_{\bar X'} > 0) +
N(\cos \theta_X \cos \theta_{\bar X'} <0)} \,,
\end{equation}
whose theoretical value derived from Eq.~(\ref{ec:dbldist}) is
\begin{equation}
A_{X \bar X'} = \frac{1}{4} C \hk_X \hk_{\bar X'} \,.
\label{ec:Ath}
\end{equation}
If CP is conserved in the decay, for charge conjugate decay channels we have
$\hk_{X'} \hk_{\bar X} = \hk_X \hk_{\bar X'}$, so the asymmetries
$A_{X' \bar X} = A_{X \bar X'}$ are equivalent. Therefore, we can sum both
channels and drop the superscripts indicating the charge, denoting the
asymmetries by $\all$, $\anl$, etc. (CP-violating effects will be
discussed in the next section).
In semileptonic top decays we can select as spin analyser the
charged lepton, which has the largest spin analysing power, or the neutrino, as
proposed in Ref.~\cite{neutrino}. In hadronic decays 
the jets corresponding to up- and down-type quarks are very difficult to
distinguish, and one possibility is to use the least energetic jet in the top
rest frame, which corresponds to the down-type quark 61\% of the time, and has a
spin analysing power $\hk_j = 0.49$ at the tree level. An equivalent possibility
is to choose the
$d$ jet by its angular distribution in the $W^-$ rest frame \cite{mahlon}.
In both hadronic and leptonic decays the $b$ ($\bar b$) quarks can be used as
well.

\begin{figure}[htb]
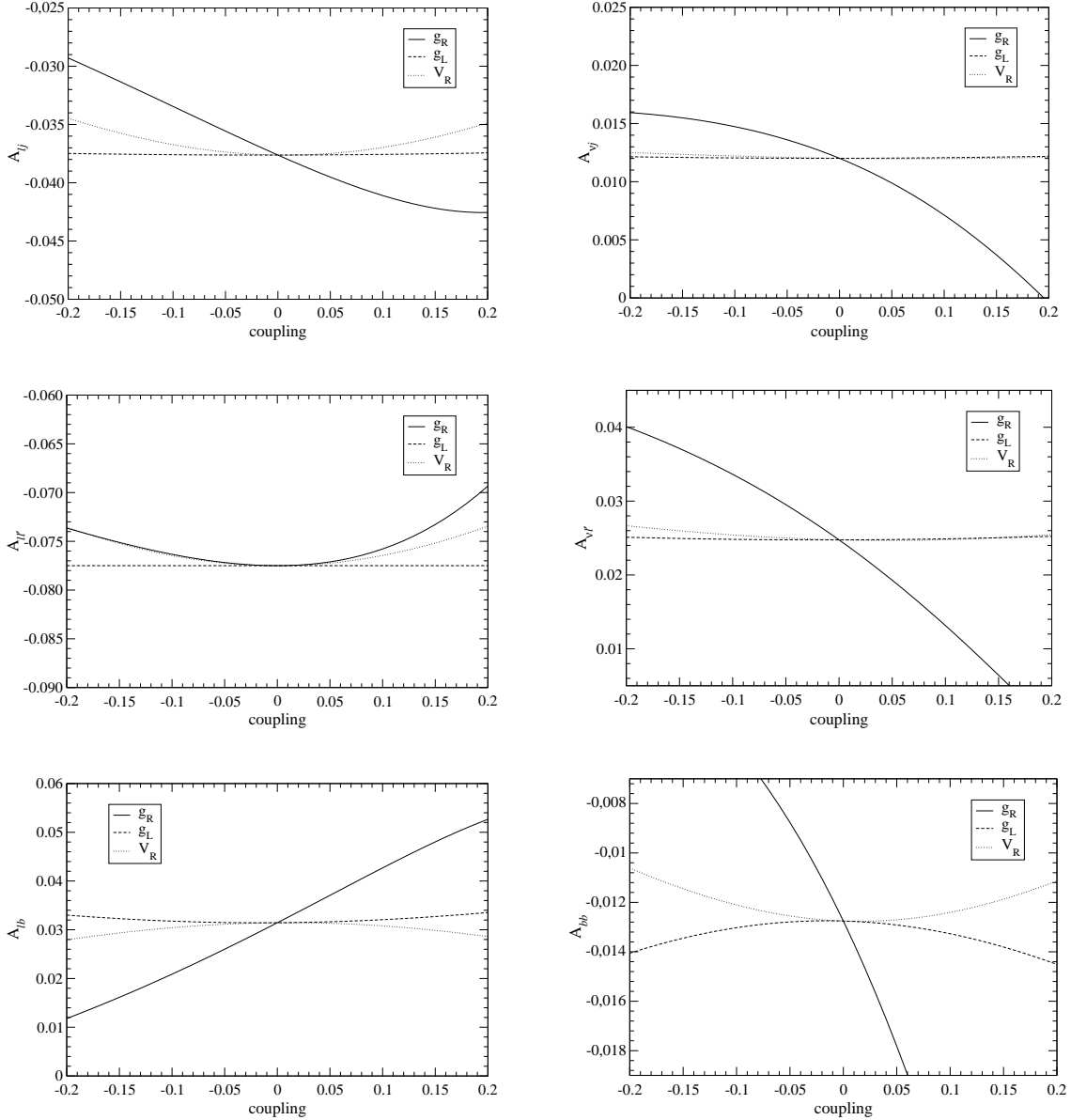

\begin{center}
\begin{tabular}{ccc}
\epsfig{file=Figs/alj.eps,height=4.8cm,clip=} & ~ &
\epsfig{file=Figs/anj.eps,height=4.8cm,clip=} \\[0.5cm]
\epsfig{file=Figs/all.eps,height=4.8cm,clip=} & ~ &
\epsfig{file=Figs/anl.eps,height=4.8cm,clip=} \\[0.5cm]
\epsfig{file=Figs/alb.eps,height=4.8cm,clip=} & ~ &
\epsfig{file=Figs/abb.eps,height=4.8cm,clip=} 
\end{tabular}
\caption{Dependence of several spin correlation asymmetries
on the couplings $\gr$, $\gl$ and $\vr$, for the CP-conserving case.}
\label{fig:aspin1}
\end{center}
\end{figure}

In the lepton $+$ jets decay mode of the $t \bar
t$ pair, $t \bar t \to \ell \nu b jj \bar b$ we choose the two
asymmetries $\alj$, $\anj$, for which we obtain the SM tree-level values 
$\alj = -0.0376$, $\anj = 0.0120$. With the precision expected
for their measurement at LHC \cite{nuestro}, the measurements
$\alj \simeq -0.0376 \pm 0.0058$, $\anj \simeq 0.0120 \pm 0.0056$ are feasible.
The dependence of these asymmetries on anomalous $Wtb$ couplings is depicted in
Fig.~\ref{fig:aspin1} (we remind the reader that the $y$ axis
scales are chosen so that the range approximately corresponds to  two standard
deviations around the theoretical SM value). These plots are obtained
using Eqs.~(\ref{ec:kappas}),(\ref{ec:Ath}). We have checked, using
high-statistics Monte Carlo simulations, that finite width effects are rather
small, so that Eqs.~(\ref{ec:kappas}),(\ref{ec:Ath}) can be used to make
accurate predictions for spin correlation asymmetries.
In the dilepton channel $t \bar t \to \ell \nu b \ell' \nu \bar b$ we select the
asymmetries $\all$, $\anl$, whose SM values are
$\all = -0.0775$, $\anl = 0.0247$. The uncertainty in their measurement can be
estimated from Refs.~\cite{marsella,nuestro}, yielding
$\all \simeq -0.0775 \pm 0.0060$, $\anl \simeq 0.0247 \pm 0.0087$. Their
variation when anomalous couplings are present is shown in
Fig.~\ref{fig:aspin1}. We also plot (in this case with arbitrary $y$ axis
scales) the asymmetries $\alb$, $\abb$, which can be
measured either in the semileptonic or dilepton channel. Their SM values are
$\alb = 0.0314$, $\abb = -0.0128$, but the experimental sensitivity has been
not estimated as yet. We expect that it
may be of the order of 10\% for $\alb$, and worse for $\abb$.

The comparison of these plots with the ones in previous sections makes apparent
that, given the experimental accuracies achievable in each case, spin
correlation asymmetries are much less sensitive to non-standard $Wtb$ couplings.
This implies that, if no deviations are found in the measurement of
the helicity ratios $\rhpm$ and angular asymmetries $A_\pm$, spin-dependent
asymmetries can be used to test
$t \bar t$ spin  correlations in the production, without contamination from
possible new interactions in the decay.
In particular, this is the case of $\all$
and $\alj$, whose relative accuracy is better, 7.7\% and 15\%, respectively.
The determination of the correlation factor $C$ in Eq.~(\ref{ec:C}) from these
asymmetries would eventually give
\begin{eqnarray}
\all & \rightarrow & C = 0.310 \pm 0.024 ~\text{(exp)} 
 ~^{+0.}_{-0.0043} ~(\delta \vr) 
 ~^{+1 \times 10^{-5}}_{-3 \times 10^{-6}} ~(\delta \gl)
 ~^{+7 \times 10^{-6}}_{-0.0004} ~(\delta \gr)
 \,, \notag \\
\alj & \rightarrow & C = 0.310 \pm 0.045 ~\text{(exp)} 
 ~^{+0.}_{-0.0068} ~(\delta \vr)
 ~^{+0.0001}_{-0.0008} ~(\delta \gl)
 ~^{+0.0004}_{-0.0009} ~(\delta \gr)
  \,.
\label{ec:Cresul}
\end{eqnarray}
The first error quoted corresponds to the experimental (systematic and
statistic) uncertainty. The other ones are theoretical uncertainties
obtained varying the anomalous couplings (one at a time). The
confidence level (CL) corresponding to the intervals quoted is 68.3\%.
The numerical comparison of the different
terms in Eqs.~(\ref{ec:Cresul}) also shows that $\alj$ and $\all$ are much less
sensitive to non-standard top couplings than $\Ap$, $\Am$ and $\rhpm$.
It must also be noted that, since all asymmetries depend on the production
mechanism through the common factor $C$, their ratios do not (to leading order),
and hence they are clean probes for anomalous couplings. The precision in the
measurement of asymmetry ratios is still to be determined, but at any rate it is
expected to be worse than for spin-independent observables discussed in the
previous sections.

It is also interesting to study the relative distribution of one spin
analyser from the $t$ quark and other from the $\bar t$. Let $\varphi_{X \bar
X'}$ be the angle between the three-momentum of $X$ (in the $t$ rest frame) and 
of $\bar X'$ (in the $\bar t$ rest frame). The angular distribution can be
written as \cite{bern}
\begin{equation}
\frac{1}{\sigma} \frac{d\sigma}{d \cos \varphi_{X \bar X'}} =
\frac{1}{2} (1+D \, \hk_X \hk_{\bar X'} \cos \varphi_{X \bar X'}) \,,
\label{ec:sgldist}
\end{equation}
with $D$ a constant defined by this equality. In our simulations we obtain the
tree-level value $D = -0.217$, while at one loop $D = -0.238$ \cite{bern},
with a theoretical uncertainty of $\sim 4$\%. 
Corresponding to these distributions, we can build the asymmetries
\begin{equation}
\tilde A_{X \bar X'} \equiv \frac{N(\cos \varphi_{X \bar X'} > 0) 
- N(\cos \varphi_{X \bar X'} < 0)}{N(\cos \varphi_{X \bar X'} > 0) +
N(\cos \varphi_{X \bar X'} < 0)}  = \frac{1}{2} D \hk_X \hk_{\bar X'} \,.
\end{equation}
For charge conjugate decay channels the distributions can be summed, since
$\hk_{X'} \hk_{\bar X} = \hk_X \hk_{\bar X'}$ provided CP is conserved in the
decay. The dependence of these asymmetries $\tilde A_{X \bar X'}$ on anomalous
couplings is (within the production $\times$ decay factorisation approximation)
exactly the same  as for the asymmetries $A_{X \bar X'}$ defined above, and
plots are not presented for brevity. Simulations are available for $\aljt$ and
$\allt$, whose theoretical SM values are $\aljt = 0.0527$, $\allt = 0.1085$. The
experimental precision expected \cite{nuestro,marsella} is
$\aljt \simeq 0.0554 \pm 0.0061$, $\allt \simeq 0.1088 \pm 0.0056$.
This is a better precision than for $\alj$ and $\alj$, respectively, 
but still not competitive in the
determination of the $Wtb$ vertex structure.\footnote{A special situation
occurs if
there is a fine-tuned cancellation between two nonzero $\vr$ and $\gl$
couplings leading to small effects in $W$ helicity fractions and related
quantities. These cancellations are possible, and in such particular case
the measurement of spin asymmetries like $\all$ and $\allt$ (which are
insensitive to $\gl$ but sensitive to $\vr$) or single top production may be
used to obtain additional information about anomalous $Wtb$ couplings.}
Instead, we can use them to test top spin correlations.
 From these asymmetries one can extract the value of
$D$, obtaining
\begin{eqnarray}
\allt & \rightarrow & D = -0.217 \pm 0.011 ~\text{(exp)} 
 ~^{+0.0031}_{-0.} ~(\delta \vr) 
 ~^{+2 \times 10^{-6}}_{-8 \times 10^{-6}} ~(\delta \gl)
 ~^{+0.0003}_{-0.} ~(\delta \gr)
 \,,  \notag \\
\aljt & \rightarrow & D = -0.217 \pm 0.024 ~\text{(exp)} 
 ~^{+0.0047}_{-0.} ~(\delta \vr) 
 ~^{+0.0006}_{-9 \times 10^{-6}} ~(\delta \gl)
 ~^{+0.0004}_{-6 \times 10^{-5}} ~(\delta \gr)
 \,. \notag \\
\label{ec:Dresul}
\end{eqnarray}
The errors quoted correspond to the experimental (systematic + statistical)
uncertainty and the variation when one of the anomalous couplings is allowed to
be nonzero. As in the previous case, the measurement of a ratio of two
asymmetries $\tilde A_{X \bar X'}$ provides a clean probe for anomalous
couplings, but with a precision expected to be worse than for spin-independent
observables.

\section{Effect of complex phases in helicity fractions and spin
asymmetries}
\label{sec:a}

\begin{figure}[htb]
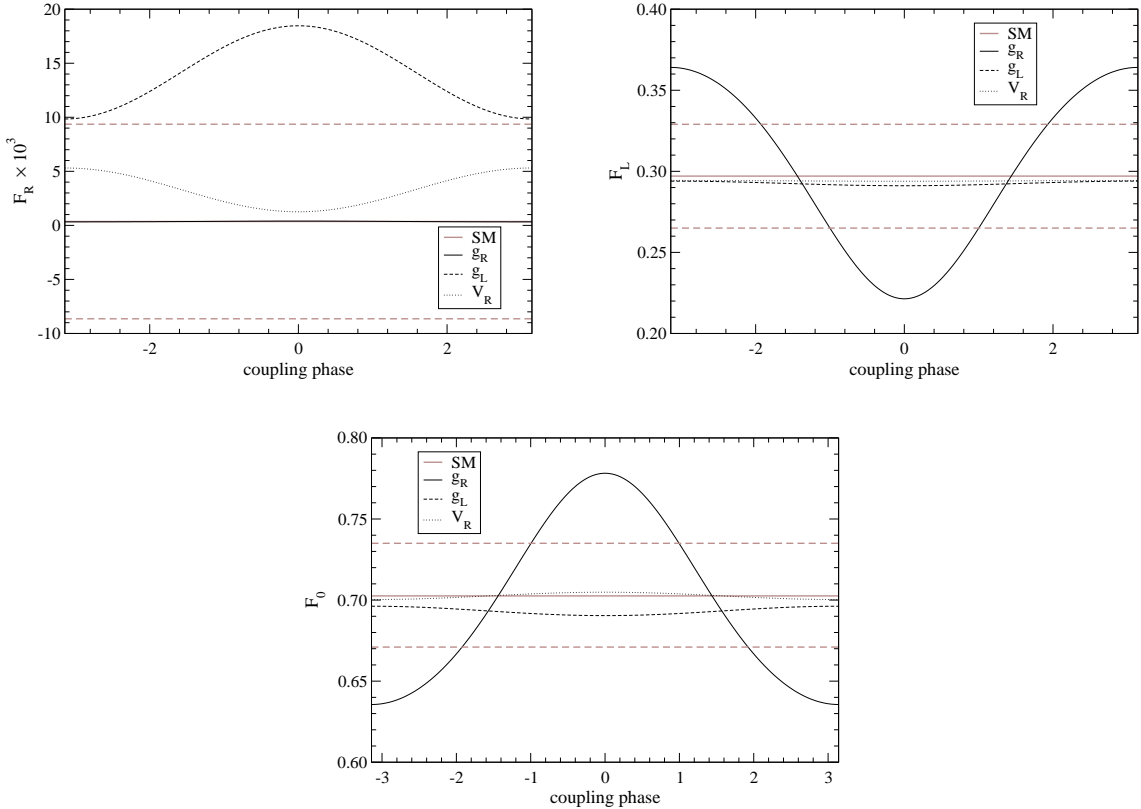

\begin{center}
\begin{tabular}{ccc}
\epsfig{file=Figs/fp_phase.eps,height=5.0cm,clip=} & ~ &
\epsfig{file=Figs/fm_phase.eps,height=5.0cm,clip=} \\[0.5cm]
\multicolumn{3}{c}{\epsfig{file=Figs/f0_phase.eps,height=5.0cm,clip=}}
\end{tabular}
\caption{Dependence of the helicity fractions $\Fi$ on the phases of the
anomalous couplings (see the text for details).}
\label{fig:f_phases}
\end{center}
\end{figure}

In the previous sections we have assumed that any non-standard $Wtb$ couplings
are real, either positive or negative. We have also pointed out that, if a
non-zero coupling $\gr$ exists, its phase has an important influence on $W$
helicity fractions and angular distributions determined by them.
Complex phases in $\vr$, $\gl$ and $\gr$ influence
the helicity fractions $\Fi$ through interference terms, which involve the
real parts of these couplings (assuming $\vl$ real). (Interference terms
are the most important ones for small values of $\vr$, $\gl$ and $\gr$, and for
the latter coupling they are unsuppressed.) The maximum and minimum effects
of anomalous couplings on $\Fi$ are obtained when they are real, negative or
positive (not necessarily in this order). We show in
Fig.~\ref{fig:f_phases} the values of the helicity fractions for fixed moduli
and arbitrary phases of the new couplings, $\vr = 0.1 \, e^{i \phi_{\vr}}$,
$\gl = 0.1 \, e^{i \phi_{\gl}}$, $\gr = 0.1 \, e^{i \phi_{\gr}}$ (one different
from zero at a time), to illustrate the effect of the phases. The plot scales
have been enlarged to cover all the range of variation of $\Fi$, and the
$2 \sigma$ expected limits have been marked with a gray
dashed line.

For $\gl$ and $\vr$ the deviations from the SM value are relatively stable under
variations of the phases, because the linear terms (which depend on the phase)
and quadratic ones (which do not) are comparable in magnitude. Thus, the
presence of a complex phase does not significantly affect the observability of
the coupling. On the other hand, for $\gr$ the effect of the phase is dominant,
and we notice that for phases $\phi_{\gr} = \pm \pi/2$ the helicity fractions
are very close to their SM values, so that
a purely imaginary coupling
$\gr \sim O(0.1)$ could remain unnoticed in an analysis of angular
distributions. We also note that
the plots are symmetric with respect to the $y$ axis, because $\Fi$
depend on $\RE\,\vr$, $\RE\,\gl$, $\RE\,\gr$ and the moduli. This
also implies that complex couplings have the same effect on the helicity
fractions in $t$ and $\bar t$ decays. Therefore, the comparison in $t$ and
$\bar t$ decays of the angular distributions studied does not
give any extra information regarding the complex phases, and
further observables are needed in order to investigate this
possibility. We have also analysed the phase dependence of the most interesting
spin asymmetries,
$\alj$, $\aljt$, $\all$ and $\allt$ (in semileptonic decays we consider
$W^+ \to \ell^+ \nu$, $W^- \to \bar q q'$). The effect of the phases is barely
detectable, even with
a greater experimental precision, and in any case the phase and modulus of an
eventual anomalous coupling measured could not be disentangled.

\section{CP-violating asymmetries}
\label{sec:6}

In this section we examine whether the existence of complex phases in the
anomalous couplings can be detected using CP-violating asymmetries.
The possibility of a relatively large imaginary coupling $\gr$ is particularly 
intriguing, since angular distributions are very sensitive to this coupling
provided its phase is not close to $\pm \pi/2$. For the other couplings,
$\gl$ and $\vr$, the situation is not so dramatic, because the observability
mainly depends on their moduli.

The spin asymmetry \cite{peskin}
\begin{equation}
A_{CP}^{RL} = \frac{\sigma(t_R \bar t_R) - \sigma(t_L \bar t_L)}
{\sigma(t_R \bar t_R) + \sigma(t_L \bar t_L)} \,,
\end{equation}
is CP-violating, and vanishes at the tree level in QCD interactions.
The top and antitop spins can be inferred using their
decay products as spin analysers, in the same way as in section \ref{sec:5}. We
thus write
\begin{equation}
A_{CP}^{RL} = \frac{N(\cos \theta_X > 0,\cos \theta_{\bar X'} > 0) -
N(\cos \theta_X < 0,\cos \theta_{\bar X'} < 0)}
{N(\cos \theta_X > 0,\cos \theta_{\bar X'} > 0) +
N(\cos \theta_X < 0,\cos \theta_{\bar X'} < 0)} \,.
\end{equation}
Even with $A_{CP}^{LR}$ vanishing in the production process, complex
phases in the decay could in principle lead to an observable asymmetry.
We have considered the dilepton channel, in which larger asymmetries are
expected because of the higher spin analysing power of the charged leptons.
(Other possibility to measure this asymmetry in the dilepton channel would be
to consider the charged lepton energies \cite{grzad3}.)
We have found that, for anomalous couplings of order $O(0.1)$ and arbitrary
phases, this asymmetry remains below the permille level, and with values
consistent with zero within Monte Carlo uncertainty. Observation of such
asymmetry would then unambiguously indicate CP-violating effects in $t
\bar t$ production, which are possible, for instance, in two Higgs doublet
models \cite{peskin,bernCP1,osland}.

\begin{figure}[htb]
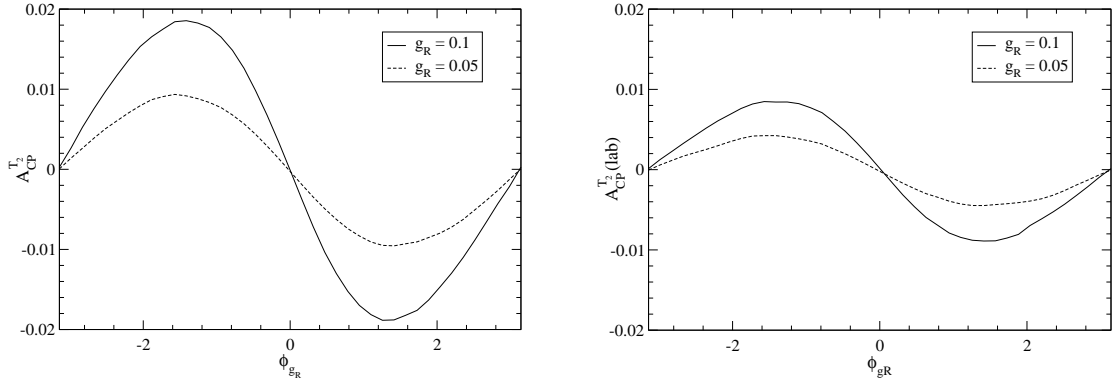

\begin{center}
\begin{tabular}{ccc}
\epsfig{file=Figs/aCP.eps,height=5.0cm,clip=} & ~ &
\epsfig{file=Figs/aCPlab.eps,height=5.0cm,clip=}
\end{tabular}
\caption{Dependence of the CP asymmetry $A_{CP}^{T_2}$, defined in top quark
rest frame (left) and laboratory system (right), on the phase of $\gr$.}
\label{fig:acp}
\end{center}
\end{figure}

We also investigate triple-product asymmetries defined in the dilepton channel,
\begin{equation}
A_{CP}^{T_i} = \frac{N(T_i > 0) - N(T_i < 0)}{N(T_i > 0) + N(T_i < 0)} \,,
\label{ec:aT}
\end{equation}
where the triple products $T_i$ are \cite{bernCP1,bernCP2}
\begin{eqnarray}
T_1 & = & \hat e \cdot (\vec p_{\ell^+} - \vec p_{\ell^-}) \;
 (\vec p_{\ell^+} \times \vec p_{\ell^-}) \cdot \hat e \,, \notag \\
T_2 & = & (\vec p_b - \vec p_{\bar b}) \cdot
  (\vec p_{\ell^+} \times \vec p_{\ell^-}) \,, \notag \\
T_3 & = & (\vec p_t - \vec p_{\bar t}) \cdot
  (\vec p_{\ell^+} \times \vec p_{\ell^-}) \,.
\end{eqnarray}
The unit vector $\hat e$ is taken in the beam direction and the particle momenta
follow obvious notation. Final state particle momenta can be measured in the
laboratory frame or, if the
kinematics of the event is completely reconstructed, in other reference system.
For asymmetries built using $T_1$ and $T_3$ we have found values $O(10^{-4})$,
and compatible with zero, taking anomalous couplings of order
$O(0.1)$ with arbitrary phases. On the other hand, we have found that
$A_{CP}^{T_2}$ is sensitive to a $\gr$ coupling of this magnitude. The asymmetry
is larger if the charged lepton and $b$ quark momenta are measured in
the respective rest frames of the decaying top quarks. However, its
observability will depend on systematic errors associated to the reconstruction,
which have not been estimated as yet, and may be better when defined in the
laboratory frame. The asymmetry $A_{CP}^{T_2}$ in both reference systems
as plotted in Fig.~\ref{fig:acp}, for couplings $\gr = 0.05$, $\gr = 0.1$ with
arbitrary phases.

Other asymmetries discussed in the literature are based on the differences
\begin{eqnarray}
\Delta_1 & = & E_{\ell^+} - E_{\ell^-} \,, \notag \\
\Delta_2 & = & \vec p_{\bar t} \cdot \vec p_{\ell^+} -  
  \vec p_t \cdot \vec p_{\ell^-} \,, \notag \\
\Delta_3 & = & \cos \theta_{\ell^+} - \cos \theta_{\ell^-} \,.
\end{eqnarray}
These quantities do not involve the product 
$(\vec p_{\ell^+} \times \vec p_{\ell^-})$, and then they can be measured in the
semileptonic channel too \cite{bernCP2}. The asymmetries are built as in
Eq.~(\ref{ec:aT}), and take values below the permille level for anomalous
couplings of the size considered in this section. For completeness, we have also
considered the P-violating asymmetry
\begin{equation}
A_P^{RL} = \frac{\sigma(t_R \bar t_L) - \sigma(t_L \bar t_R)}
{\sigma(t_R \bar t_L) + \sigma(t_L \bar t_R)} \,,
\end{equation}
which is measured using the charged leptons as spin analysers,
\begin{equation}
A_P^{RL} = \frac{N(\cos \theta_X > 0,\cos \theta_{\bar X'} < 0) -
N(\cos \theta_X < 0,\cos \theta_{\bar X'} > 0)}
{N(\cos \theta_X > 0,\cos \theta_{\bar X'} < 0) +
N(\cos \theta_X < 0,\cos \theta_{\bar X'} > 0)} \,.
\end{equation}
We have found $A_P^{RL} \sim 1.5 \times 10^{-3}$, and insensitive to anomalous
couplings $O(0.1)$. This asymmetry can be sizeable in SM extensions
\cite{PV}.

We finally emphasise that, even in the cases where they are insensitive to
complex anomalous $Wtb$ couplings, the CP-violating asymmetries studied here
are still very useful to
disentangle CP violation in the production and the decay. If these asymmetries
are found to be non-vanishing, they clearly
signal CP violation in $t \bar t$ production. On the other hand, the imaginary
part of $\gr$ can be probed using $A_{CP}^{T_2}$.

\section{Summary}
\label{sec:7}

New physics, if it exists close to the electroweak scale, may manifest itself
through non-standard top interactions. In this paper we have discussed top pair
decays at LHC as a probe of the $Wtb$ vertex. We have examined angular and
energy distributions, as well as asymmetries, involving or not the top quark
polarisation. Among the observables discussed, the best sensitivity to anomalous
$Wtb$ couplings is given by the $W$ helicity fractions $\Fi = \Gamma_i /
\Gamma$, $i=L,R,0$, and related observables. We have obtained analytical
expressions for $\Fi$, for a general CP-violating $Wtb$ vertex with the top
quark and $W$ boson on their mass shell, and keeping a non-zero bottom quark
mass. We have shown, comparing with exact numerical results, the high accuracy
of this approximation when studying angular distributions. We have also pointed
out the importance of keeping the bottom mass in the calculations, in contrast
with previous studies in the literature.

$W$ helicity fractions can be extracted from a fit to the charged lepton angular
distribution in $W$ rest frame. The same analysis can be used to determine the
helicity ratios $\rhpm = \Gpm / \Gz$, which are actually more
sensitive to $\vr$ and $\gl$-type anomalous couplings, given the experimental
uncertainties (dominated by systematics already for a luminosity of
10 fb$^{-1}$)
associated to each observable. A simpler method to probe the $Wtb$ vertex,
without the need of a fit to the charged lepton distribution, is through angular
asymmetries. We have introduced two new asymmetries $\Ap$ and $\Am$, in addition
to the $\ell W$ (or $\ell b$) forward-backward asymmetry $\afb$ previously
studied \cite{prd}. These new asymmetries allow us to: (i) obtain more precise
bounds on
anomalous couplings than $\afb$, comparable with those obtained from $\rhpm$
and $\Fi$, and even better for a $\gr$ coupling;
(ii) determine the helicity fractions with a fair accuracy without fitting the
charged lepton distribution.
The $W$ helicity fractions determine the charged lepton energy distribution in
top rest frame as well. Energy asymmetries can be built, but they are less
suited for the study of anomalous couplings because the approximation of
considering the top quark and $W$ boson on shell is worse, and experimental
uncertainties on energy asymmetries are larger. The best limits found, using
single measurements, are
\begin{align}
& -0.029 \leq \vr \leq 0.099 \quad (\rhp) \,, \notag \\
& -0.046 \leq \gl \leq 0.013 \quad (\rhp) \,, \notag \\
& -0.019 \leq \gr \leq 0.018 \quad (\Ap) \,.
\label{ec:best}
\end{align}
Limits can be improved by combining the measurements of $A_\pm$ and $\rhpm$.
The theoretical predictions for these and other observables have been
implemented in a computer program {\tt TopFit}, which allows to extract
combined limits on anomalous couplings from a given set
of observables, following the statistical approach outlined in appendix
\ref{sec:c}. Detailed results including the correlation of the various
observables (computed from Monte Carlo simulations) are beyond the
scope of this paper, and have been
presented elsewhere \cite{nuestro2}.

Spin correlations and spin-dependent asymmetries probe not only the $Wtb$
interactions but also the dynamics of $t \bar t$ production. Their study is very
interesting from a theoretical point of view, because they are sensitive to {\em
e.g.} the exchange of a scalar particle in $s$ channel \cite{bernhiggs} or
anomalous $gtt$ couplings \cite{cheung}. It is then
crucial to disentangle new physics in the production from possible
anomalous $Wtb$ couplings. This could be done, for instance, considering ratios
of spin correlation asymmetries or, even better, using the strict bounds on
anomalous couplings obtained from top decay angular distributions.

The dependence of spin correlation asymmetries on $Wtb$ anomalous couplings
occurs through the ``spin analysing power'' constants of top quark decay
products. We have calculated these constants for a general CP-conserving $Wtb$
vertex, in the narrow width approximation. It has been shown that the
sensitivity of spin correlation asymmetries to top anomalous couplings is much
weaker than for helicity fractions and related observables. Then, we have set
explicit limits on the variation
of two factors $C$, $D$ (which measure the $t \bar t$ spin correlation) due to
possible anomalous couplings not detected in other processes, {\em i.e.} within
the ranges in Eqs.~(\ref{ec:best}). The possible variation in $C$, $D$ is
much smaller than the experimental precision expected for their measurement,
\begin{align}
C & = 0.310 \pm 0.024 ~\text{(exp)} 
 ~^{+0.}_{-0.0043} ~(\delta \vr) 
 ~^{+1 \times 10^{-5}}_{-3 \times 10^{-6}} ~(\delta \gl)
 ~^{+7 \times 10^{-6}}_{-0.0004} ~(\delta \gr)
 \,, \notag \\
D & = - 0.217 \pm 0.011 ~\text{(exp)} 
 ~^{+0.0047}_{-0.} ~(\delta \vr) 
 ~^{+0.0006}_{-9 \times 10^{-6}} ~(\delta \gl)
 ~^{+0.0004}_{-6 \times 10^{-5}} ~(\delta \gr)
 \,.
\end{align}
Hence, any deviation observed experimentally should correspond to new physics
in the production. On the other hand, in ratios of two spin asymmetries 
$A_{X \bar X'}$ ($\tilde A_{X \bar X'}$) the common factors $C$ ($D$) cancel,
and thus the ratios can cleanly probe non-standard top couplings. These
observables have also been implemented in the computer program {\tt TopFit}, and
estimates for their expected precision will be presented elsewhere.

Finally, we have addressed the possibility of complex anomalous $Wtb$ couplings
$\gl$, $\gr$, $\vr$. Complex phases in these terms influence helicity fractions
and related quantities via the interference with the dominant SM coupling $\vl$
(which we have normalised to unity). For $\vr$ and $\gl$, quadratic and
interference
terms have the same magnitude, and the effect of phases is not very relevant.
For $\gr$, however, the interference term dominates, and the dependence on the
phase is very strong. One finds that a $\gr$ coupling with a phase close to
$\pm \pi/2$ has little effect on angular distributions, and even with a
relatively large modulus it could remain unnoticed in such analyses. The same
has been found for spin correlation asymmetries. However,
we have shown that a CP asymmetry based on the triple product
$(\vec p_b - \vec p_{\bar b}) \cdot (\vec p_{\ell^+} \times \vec p_{\ell^-})$ is
sensitive to a complex $\gr$, taking values up to $\pm 2$\% for
$\gr = \mp 0.1 i$.
If this asymmetry can be measured at LHC with a precision below the percent
level, it could help to measure or bound $\gr$. The remaining CP
asymmetries analysed are very small, and insensitive to anomalous couplings of
this size. Therefore, they can be used to isolate CP violating
effects in $t \bar t$ production \cite{bernCP1,osland}.
On the other hand, single top production at LHC can probe the $Wtb$
interaction, and $B$ or super-$B$ factories, with precise measurements
of CP asymmetries {\em e.g.} in $b \to s \gamma$, might also give
indirect evidence for (real or complex) anomalous $Wtb$ couplings, helping to
determine the structure of this vertex.

\vspace{1cm}
\noindent
{\Large \bf Acknowledgements}

The work of J.A.A.-S. has been supported by a MEC Ramon y Cajal contract and
project FPA2003--09298--C02--01, and by Junta de Andaluc\'{\i}a
through project FQM-101. The
work of J.C., N.C. (grant SFRH/BD/13936/2003), A.O. and F.V.
(grant SFRH/BD/ 18762/2004) has been supported by
Fundaç\~ao para a Ci\^encia e a Tecnologia.

\appendix

\section{Effect of $\boldsymbol{m_b}$ in the helicity fractions}
\label{sec:b}

\begin{figure}[htb]
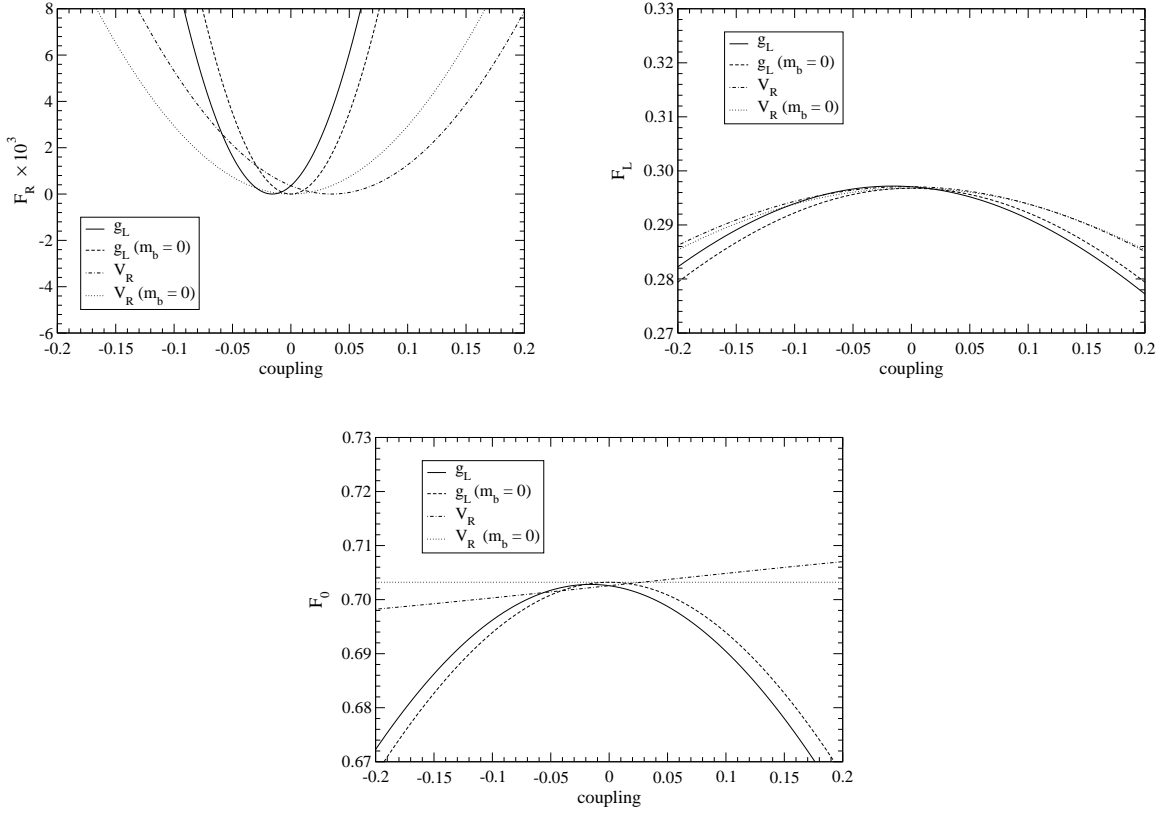

\begin{center}
\begin{tabular}{ccc}
\epsfig{file=Figs/fp_mb0.eps,height=5.0cm,clip=} & ~ &
\epsfig{file=Figs/fm_mb0.eps,height=5.0cm,clip=} \\[0.5cm]
\multicolumn{3}{c}{\epsfig{file=Figs/f0_mb0.eps,height=5.0cm,clip=}}
\end{tabular}
\caption{Dependence of the helicity fractions $\Fi$
on $\vr$ and $\gl$, for $m_b = 4.8$ GeV and neglecting the $b$ quark mass.}
\label{fig:mbeff}
\end{center}
\end{figure}

As it can be observed in Eqs.~(\ref{ec:gammas}), interference terms
involving $\vr$ (or $\gl$) and the dominant SM coupling $\vl$ are proportional
to $x_b = m_b/m_t$. These terms are of equal size as the quadratic terms for
small $\vr$, $\gl$, and cannot be neglected in the analysis. To illustrate their
importance, we plot in Fig.~\ref{fig:mbeff} the dependence of the three helicity
fractions on the anomalous couplings, for $m_b = 4.8$ GeV and neglecting $m_b$.
The differences are apparent for $\fp$, and for $\fz$ we have the extreme
situation that the only dependence of this quantity on $\vr$ is through the
$x_b$ term.

The $m_b$ dependence of the limits generates a small uncertainty due to the
uncertainty in $m_b$, for which we use the $b$ quark pole mass. This $b$ mass
definition has an ambiguity of the order of $\lqcd \simeq 220$
MeV (for other definitions the uncertainty is smaller). The variation of the
limits in Eq.~(\ref{ec:best}) when $m_b$ is taken
as 4.8 GeV $\pm \lqcd$ is presented in Table~\ref{tab:sysmb} (we
display additional digits in order to better illustrate the variation). The
effect is larger for $\vr$ and $\gl$, as it is expected from the discussion in
section \ref{sec:2}. Nevertheless, the uncertainty only amounts to a few
percent.

\begin{table}[htb]
\begin{center}
\begin{tabular}{ccccccc}
& \multicolumn{3}{c}{Lower limit} & \multicolumn{3}{c}{Upper limit} \\
Coupling & $m_b - \lqcd$ & $m_b$ & $m_b + \lqcd$ & $m_b - \lqcd$ & $m_b$ &
$m_b + \lqcd$ \\
$\vr$ & -0.0303 & -0.0293 & -0.0281 & 0.0977 & 0.0994 & 0.1077 \\
$\gl$ & -0.0444 & -0.0456 & -0.0461 & 0.0140 & 0.0135 & 0.0127 \\
$\gr$ & -0.0192 & -0.0194 & -0.0192 & 0.0181 & 0.0180 & 0.0178 
\end{tabular}
\end{center}
\caption{Influence of $m_b$ in the limits in Eqs.~(\ref{ec:best}): variation
when the $b$ quark mass is taken at its central value $m_b = 4.8$ GeV, or adding
and subtracting a small uncertainty $\lqcd = 220$ MeV.}
\label{tab:sysmb}
\end{table}

\newpage
\section{Extraction of limits from observables}
\label{sec:c}

The derivation of limits on the anomalous couplings from the measurement of
the experimental observables discussed has to be done with special care, due to
the non-linear dependence of the
latter on the former. In this appendix we explain the method we have used
to obtain our limits.

Let us denote by $O$ a generic observable, {\em e.g.} an angular asymmetry,
and $x$ an unknown parameter (in our case an anomalous coupling) upon which this
observable depends, and for which we want to obtain a confidence interval.
$O$ is experimentally measured and is assumed to obey a
Gaussian distribution (with mean and standard deviation given by its
measurement). However, if the dependence $O(x)$ is non-linear in the
region of interest, the probability density function (p.d.f.) derived for the 
parameter $x$ will no longer be a Gaussian and a Monte Carlo method must be used
to determine a confidence interval on $x$.

We determine the p.d.f. of $x$ numerically, using the acceptance-rejection
method: we iteratively
(i) generate a random value (with uniform probability) $x_i$ within a suitable
interval; (ii) evaluate the probability of $O(x_i)$, given by the p.d.f. of $O$;
(iii) generate an independent
random number $r_i$ (with uniform probability); and (iv) accept the value $x_i$
if the probability of $O(x_i)$ is larger than $r_i$. The resulting set of values
$\{x_i\}$ is distributed according to the p.d.f. of $x$ given by the measurement
of $O$. The determination of a
central interval with a given confidence level (CL) $\gamma$ is done
numerically, requiring: (a) that it contains a fraction $\gamma$ of the total
number of values $\{x_i\}$; (b) that is central, {\em i.e.} fractions
$(1-\gamma)/2$ of the values generated are on each side of the interval.

We have applied this method to obtain the limits on Tables~\ref{tab:limits1}
and \ref{tab:limits2}, keeping only one of the couplings non-vanishing at a
time. We point out that:
\begin{enumerate}
\item The dependence on $\gr$ of the observables $\Fi$, $\rhpm$, $A_\pm$ and 
$\afb$ is approximately linear, as it can be observed in
Figs.~\ref{fig:widths}--\ref{fig:asim}. Therefore, the limits on this coupling
can be approximately obtained directly
from these plots using the method in
Refs.~\cite{marsella,nuestro}: for a given observable
$O$, intersecting the plot of $O(\gr)$ with the two horizontal lines
$O = O_\text{exp} \pm \Delta O$, which correspond to the $1 \sigma$ variation
of $O$, gives the $1 \sigma$ interval (with a 68.3\% CL) on $\gr$.

\item The dependence on $\gl$ and $\vl$ is highly non-linear (the region of
interest is at the extreme of a quadratic function), and appreciable
differences are found between the Monte Carlo and the intersection methods. For
example, the ``$1 \sigma$'' limit on $\vr$ obtained with the intersection method
from the (hypothetical) measurement $\rhp \simeq 0.0005 \pm 0.0026$ is
$-0.051 \leq \vr \leq 0.12$. However,
this interval has a confidence level of 85.6\%, and the true 68.3\% central
interval obtained from the same measurement with the Monte Carlo
method outlined above is $-0.029 \leq \vr \leq 0.099$.
Although overcoverage is not as bad as undercoverage, it is quite desirable that
confidence intervals have exactly the CL they are supposed to
have.

\end{enumerate}

A similar procedure is applied to estimate the
theoretical uncertainties in Eqs.~(\ref{ec:Cresul}) and (\ref{ec:Dresul}) due to
possible anomalous couplings.

\end{document}